\documentclass[paper]{JHEP3}

\usepackage[dvips]{graphicx}
\usepackage{epsfig}
\usepackage{amssymb}
\usepackage{amsmath}
\usepackage{cite}
\usepackage{axodraw4j}
\usepackage{caption}

\def\v{\begingroup\obeyspaces\u}
\def\u#1{\tt#1\endgroup}

\preprint{Cavendish--HEP--09/20}

\title{Double Parton Distributions Incorporating Perturbative QCD Evolution and 
Momentum and Quark Number Sum Rules}

\author{Jonathan R.\ Gaunt$^1$ and W. James\ Stirling$^{1,2}$\\
  $^1$Cavendish Laboratory, University of Cambridge,
  J.J.\ Thomson Avenue, Cambridge CB3 0HE, U.K.\\
$^2$ Department of Physics and Institute for Particle Physics Phenomenology, University of Durham, DH1 3LE, U.K.}

\abstract{It is anticipated that hard double parton scatterings will occur
frequently in the collisions of the LHC, producing interesting signals and
significant backgrounds to certain single scattering processes. 
For double scattering processes in which the
same hard scale $t = \ln(Q^2)$ is involved in both collisions, we require the
double parton distributions (dPDFs) $D_h^{j_1j_2}(x_1,x_2;t)$  in order to
make theoretical predictions of their rates and properties. We describe the development of
a new set of leading order dPDFs that represents an improvement on approaches used previously.
First, we derive momentum and number sum rules that the dPDFs must satisfy. The fact that
these must be obeyed at any scale is used to construct improved dPDFs at the input scale
$Q_0$, for a particular choice of input scale ($Q_0^2 = {1}$~GeV$^2$) and corresponding
single PDFs (the MSTW2008LO set). We then describe a novel program which uses a direct
$x-$space method to numerically integrate the LO DGLAP equation for the dPDFs, and which 
may be used to evolve the input dPDFs to any other scale. This program has been used along 
with the improved input dPDFs to produce a set of publicly available dPDF grids covering 
the ranges $10^{-6}<x_1<1$, $10^{-6}<x_2<1$, and $1<Q^2<10^9\rm{~GeV}^2$.}


\begin{document}

\pagebreak

\section{Introduction} \label{sec:intro}

In the standard framework for calculating inclusive cross sections for hard
scattering processes in hadron-hadron collisions, it is assumed that only one
hard interaction occurs per collision (plus multiple soft interactions). This
assumption is typically justified on the grounds that the probability of a hard
parton-parton interaction in a collision is very small. Thus the probability of
having two or more hard interactions in a collision is highly suppressed with
respect to the single interaction probability.

Hadron-hadron collisions in which two (or more) distinct pairs of partons hard
scatter are nevertheless possible. Early theoretical studies of double scattering
were carried out in the context of the parton model \cite{Landshoff:1978fq, Takagi:1979wn ,Goebel:1979mi}
, with subsequent extension to perturbative QCD \cite{Paver:1982yp, Humpert:1983pw, Mekhfi:1983az,
Humpert:1984ay, Ametller:1985tp, Mekhfi:1985dv, Halzen:1986ue, Sjostrand:1987su, Mangano:1988sq, 
Godbole:1989ti, Drees:1996rw, Calucci:1997uw, Calucci:1999yz}
. Such processes have in fact been observed experimentally -- both in
$\sqrt{s} = 63$~GeV $pp$ collisions by the AFS collaboration at the CERN ISR
\cite{Akesson:1986iv} and more recently in $\sqrt{s} = 1.8$~TeV $p\bar{p}$ collisions by the
CDF collaboration \cite{Abe:1997xk} and $\sqrt{s} = 1.96$~TeV $p\bar{p}$ collisions by the
D0 collaboration \cite{Abazov:2009gc} at the Fermilab Tevatron.

The much greater energy and higher luminosity of the LHC implies that we will
observe a greater rate of events containing multiple hard interactions in this
experiment than in either of the two mentioned above. Moreover, a number of
calculations \cite{DelFabbro:1999tf, DelFabbro:2002pw, Hussein:2006xr,
Hussein:2007gj} suggest that the products from multiple interactions will
represent an important background to signals from the Higgs and other
interesting processes. Further calculations \cite{Kulesza:1999zh, Maina:2009vx, Maina:2009sj}
indicate that certain types of multiple interactions will have distinctive
signatures at the LHC, facilitating a detailed study of this process by the
experiment.

The importance of multiple scattering signals and backgrounds at the LHC
necessitates a good quantitative understanding of these processes. In
particular, it is important to understand double scattering, which will be the
dominant multiple scattering mode at the LHC.  Assuming only the factorisation
of the two hard subprocesses A and B, the cross section for this process in
proton-proton scattering may be written as:
\begin{eqnarray}
\sigma^D_{(A,B)} = \dfrac{m}{2}\sum_{i,j,k,l}\int\Gamma_{ij}(x_1,x_2,b;t_1,t_2)
\hat{\sigma}^A_{ik}(x_1,x_1')\hat{\sigma}^B_{jl}(x_2,x_2') \\ \nonumber
\times \Gamma_{kl}(x_1',x_2',b;t_1,t_2)
dx_1dx_2dx_1'dx_2'd^2b
\end{eqnarray}

The $\hat{\sigma}$s are the parton-level subprocess cross sections. These are
also encountered in single scattering, and are known for essentially all 
processes of phenomenological interest.
 The quantity $m$ is a symmetry factor that equals $1$ if $A=B$ and $2$
otherwise. The $\Gamma_{ij}(x_1,x_2,b;t_1,t_2)$ represent generalised double
distributions. They may be loosely interpreted as the inclusive probability
distributions to find a parton $i$ with longitudinal momentum fraction $x_1$ at
scale $t_1\equiv \ln(Q_1^2)$ in the proton, in addition to a parton $j$ with
longitudinal momentum fraction $x_2$ at scale $t_2\equiv \ln(Q_2^2)$, with the
two partons separated by a transverse distance $b$. The scale $t_1$ is given by
the characteristic scale of subprocess A, whilst $t_2$ is equal to the
characteristic scale of subprocess B. Note that CDF and D0 have measured double scattering 
via the $\gamma + 3$jet final state, with A corresponding to $\gamma +$jet production and
B to dijet production.

It is typically assumed that $\Gamma_{ij}(x_1,x_2,b;t_1,t_2)$ may be decomposed
in terms of longitudinal and transverse components as follows:
\begin{equation}
\Gamma_{ij}(x_1,x_2,b;t_1,t_2) = D^{ij}_h(x_1,x_2;t_1,t_2)F^i_j(b)
\end{equation}

The function $D^{ij}_h(x_1,x_2;t_1,t_2)$ has a rigorous interpretation in leading
order (LO) perturbative QCD as the inclusive probability of finding a parton $i$ with momentum
fraction $x_1$ at scale $t_1$ and a parton $j$ with  momentum fraction $x_2$ at
scale $t_2$ in the proton. Accurate prediction of double parton scattering
cross sections and event signatures requires good modelling of
$D^{ij}_h(x_1,x_2;t_1,t_2)$ and $F^i_j(b)$. In particular, one must correctly
take account of the effects of correlations in both longitudinal momenta and
transverse positions in these functions.

Correlations between the partons in transverse space are highly significant --
at the very least, they must tie the two partons together within the same
hadron. As one might suspect, their precise calculation is not possible using
perturbation theory. Existing models typically use Gaussian or exponential
forms to describe the $F^i_j(b)$, or a sum of Gaussian/exponential terms
\cite{Sjostrand:2004pf, Bahr:2008dy}.

On the other hand, correlations in longitudinal momenta are typically ignored.
The usual assumption (applied in the phenomenological calculations of
\cite{DelFabbro:1999tf, DelFabbro:2002pw, Hussein:2006xr, Hussein:2007gj,
Kulesza:1999zh, Maina:2009vx, Maina:2009sj}) is that at least for small $x_i$ values the
longitudinal momenta correlations are small, and therefore 
 $D^{ij}_h(x_1,x_2;t_1,t_2)$ may be taken
to be equal to a product of the relevant single parton distribution functions
(sPDFs) -- i.e. $D^{ij}_h(x_1,x_2;t_1,t_2) = D^{i}_h(x_1;t_1)D^{j}_h(x_2;t_2)$. 
With this assumption, plus the assumption that $F^i_j(b)$ is the same
for all parton pairs $ij$ involved in the double scattering process of
interest, the cross section $\sigma^D_{(A,B)}$ has the particularly simple
form:
\begin{eqnarray}
\sigma^D_{(A,B)} &=&\dfrac{m}{2}
\dfrac{\sigma^S_{(A)}\sigma^S_{(B)}}{\sigma_{\rm eff}}
\\ \nonumber
\sigma_{\rm eff} &=& \left[\int d^2b (F(b))^2\right]^{-1}
\end{eqnarray}
The quantity $\sigma^S_{(X)}$ is the single scattering cross section for hard
process X. The factor  $ \sigma_{\rm eff}$ in the denominator has the 
dimensions of a cross section. It can be understood as follows. Given that one hard scattering occurs,
the probability of the other hard scattering is proportional to the flux of accompanying partons; these
are confined to the colliding protons, and therefore their flux should be  inversely proportional to the area 
(cross section) of a proton. Interestingly, the CDF and D0 measurements give $\sigma_{\rm eff} \sim 15$~mb, which is 
roughly $20\%$ of the total (elastic $+$ inelastic) $p\bar{p}$ cross section at the Tevatron collider energy.

It is argued that the approximation of the $D^{ij}_h(x_1,x_2;t_1,t_2)$ as a
product of single PDFs should be particularly applicable in collider experiments where small
$x$ values are probed (i.e. large total system centre of mass energy with
respect to subprocess energy) due to the large population of partons at these
$x$ values. The CDF experimental data also agree with this assumption, with no
sign of $x$-dependence in their measured $\sigma_{\rm eff}$ over the $x$ ranges
accessible to them ($0.01-0.40$ for their first subprocess, and $0.002-0.20$
for the other) \cite{Abe:1997bp}. The D0 data confirm this result, with no measured variation
of  $\sigma_{\rm eff}$ with the (second highest) jet transverse momentum.

Even if the factorisation assumption holds at the CDF scale (i.e. $Q^2 \sim
100 - 1000$~GeV$^2$), it is unlikely that it will hold at higher scales such as will be
encountered at the LHC. In \cite{Kirschner:1979im, Shelest:1982dg}, the
behaviour of the distributions $D^{ij}_h(x_1,x_2;t) \equiv
D^{ij}_h(x_1,x_2;t,t)$ with the two hard scales set equal (hereafter known as
the dPDFs) were investigated. The authors derived an equation dictating the
scaling violations (i.e. $t$ dependence) of the dPDFs. This equation is an analogue of the  DGLAP
equation for sPDFs (sDGLAP equation) \cite{Dokshitzer:1977sg, Gribov:1972ri,
Lipatov:1974qm, Altarelli:1977zs}. An important prediction of this equation is
that, even if the dPDFs factorise at some scale $t_0$, then at {\em any}
different scale factorisation will be violated \cite{Snigirev:2003cq}. In other words, the naive
product $D_h^{i}(x_1;t)D_h^{j}(x_2;t)$ where the $D_h^{i}$s satisfy sDGLAP is {\em not} a solution
of the dDGLAP equations. Explicit
numerical solutions of the LO `double DGLAP' (dDGLAP) equation based on
factorised inputs at $Q_0^2 \sim 1\rm{~GeV}^2$ suggest that the violations are
significant even for small $x$, with deviations on the order of $10-30\%$ at
$x_1 = x_2 \sim 0.1, Q^2 \sim 10^4$~GeV$^2$ \cite{Cattaruzza:2005nu,
Korotkikh:2004bz}.

On a more fundamental level, the factorised approach is inadequate in that it fails
to take account of even very basic correlations associated with the fact that finding a
quark of given flavour reduces the chances of finding another with the same flavour. It also
neglects correlations associated with the fact that finding a parton with momentum fraction 
$x$ reduces the probability of finding another parton with momentum fraction near $1-x$ 
(apart from a crude cut-off in the probability when the sum of the two parton momenta exceeds 1). In 
other words, factorised forms do not obey the relevant momentum and number sum rules (see
section \ref{sec:sumrules}). In \cite{Sjostrand:2004pf}, a method to construct double 
(and multiple) parton distributions is suggested which is intended to capture some of the main 
features of these flavour and momentum correlations. However, this method does not introduce 
the correlations fully rigorously using the sum rules. Furthermore, the dPDFs are constructed
entirely out of sPDFs, and no use is made of the dDGLAP equation in this paper.

For the purposes of the accurate prediction of double parton scattering cross sections and signals at the LHC, there is a need for a more theoretically sound set of double distributions than either the naive factorised forms traditionally used, or the improved forms suggested in \cite{Sjostrand:2004pf}. Here, we have attempted to address this issue for the specific case of the dPDFs (with $t_1=t_2$). First, we derive sum rules corresponding to momentum and valence quark number conservation that the dPDFs must satisfy. These are used as an aid to construct `improved' dPDFs at the scale $Q_0=1$~GeV that correspond to the MSTW2008LO sPDF inputs \cite{Martin:2009iq}. The low scale dPDFs are then used as an input in a program we have written which numerically integrates the LO dDGLAP equation to higher scales. The end products of this paper are a set of LO dPDF grids covering the ranges $10^{-6}<x_1<1$, $10^{-6}<x_2<1$, $1<Q^2<10^9$~GeV$^2$, and all possibilities for the parton indices $i$ and $j$. These grids, in addition to a simple interpolation subroutine designed to extract from the grids a dPDF value at a given $x_1,x_2$ and $Q$, can be found at Ref. \cite{HepForgePage}.

This paper is organised as follows. We begin with a brief review of the dDGLAP equation in Section \ref{sec:theory}. The dPDF sum rules are introduced and discussed in Section \ref{sec:sumrules}, where we also explain how we have used these rules to construct input dPDFs at $Q_0=1$~GeV corresponding to the MSTW2008LO sPDF inputs. In Section \ref{sec:method}, the details of the numerical procedure designed to evolve the input distributions to higher scales using the LO dDGLAP equation are given. Section \ref{sec:properties} examines the ways in which our dPDFs differ from those obtained using previous approaches. Finally, we conclude in Section \ref{sec:conclusion} with a summary and discussion of potential future directions for the work.

\section{The Double DGLAP Equation} \label{sec:theory}

It is well established that in QCD, the parton content of the proton that is
observed by a hard probe with virtuality $Q^2$ (or more than one hard probe
with this virtuality) is dependent on the size of the virtuality. This
dependence is explained by the fact that a harder probe (with a shorter
associated wavelength) is able to `see' finer scale structure in the proton,
and in particular is able to resolve parton splittings that were unresolvable
using a lower $t \equiv \ln(Q^2)$ probe. This implies that parton distributions
must be dependent on the scale $t$ at which the proton is probed. There is a
shift of these distributions towards smaller $x$ values as $t$ increases, as a
consequence of a greater number of splittings being resolved.

One can visualise the change in parton distributions as one probes at steadily
higher scales $t$ than the low scale $t_0$ as a spacelike branching process
originating from the initial distributions at scale $t_0$. As one probes to
higher scales, one effectively progresses further in the branching process.

 The dDGLAP equation is a renormalisation group equation describing the change
of the dPDFs with the hard scale $t$. It is based on the leading logarithm
approximation (LLA) of perturbative QCD (the same is true for the sDGLAP
equation). This approximation corresponds to a picture of the spacelike parton
branching process from the low scale $t_0$ to the probe scale $t$ in which
gluon emissions along the parton branches are strongly ordered in transverse
momentum. Gluons emitted `earlier' in the branching process are restricted to
have smaller transverse momenta than those emitted closer to the probe scale.
The dDGLAP equation effectively resums leading powers of $[\alpha_s t]^n$
generated by these gluon emissions to give the dPDFs at scale $t$.

In \cite{Kirschner:1979im, Shelest:1982dg}, the following form for the dDGLAP
equation is derived:
\begin{align} \label{dbDGLAP}
\dfrac{dD^{{j_1}{j_2}}_h(x_1,x_2;t)}{dt} = \dfrac{\alpha_s(t)}{2\pi} \Biggl[
\sum_{j'_1}\int_{x_1}^{1-x_2}\dfrac{dx'_1}{x'_1}D^{{j'_1}{j_2}}_h(x'_1,x_2;t)P_{j'_1
\to j_1}\left(\dfrac{x_1}{x'_1}\right)
\nonumber\\
+\sum_{j'_2}\int_{x_2}^{1-x_1}\dfrac{dx'_2}{x'_2}D^{{j_1}{j'_2}}_h(x_1,x'_2;t)P_{j'_2
\to j_2}\left(\dfrac{x_2}{x'_2}\right)
\nonumber\\
+\sum_{j'}D^{j'}_h(x_1+x_2;t)\dfrac{1}{x_1+x_2}P_{j' \to j_1
j_2}\left(\dfrac{x_1}{x_1+x_2}\right) \Biggr]
\end{align}

Technically, the argument $t$ in the above is the factorisation scale (which in 
practical calculations is typically set equal to the characteristic hard scale
of the subprocesses). The renormalisation scale has been set equal to the 
factorisation scale to obtain this equation (as is conventional in a leading 
order analysis).

In addition to the dPDFs and sPDFs $D^{j}_h(x;t)$, the equation \eqref{dbDGLAP}
contains two different types of splitting functions. The first are the
well-known splitting functions $P_{i \to j}(x)$ previously encountered in the
context of the sDGLAP equation. They are given to both LO and NLO in
\cite{Ellis:1996ws}. At leading
order, the function $P_{i \to j}(x)$ may be interpreted as the probability of a
parton $i$ splitting to give a parton $j$ with a fraction $x$ of the
longitudinal momentum of the parent parton and a transverse momentum squared
much smaller than $Q^2$ (where $t \equiv \ln(Q^2)$) \cite{Altarelli:1977zs}.
The second, the $P_{i \to jk}(x)$, are new. They may be interpreted at LO as the probability of a
parton $i$ splitting to give the two partons $j$ and $k$, the first of which
has a fraction $x$ of the linear momentum of the parent parton, the second of
which has the remainder of the linear momentum $1-x$, and both of which have
transverse momentum squared much less than $Q^2$.

The splitting functions $P_{i \to i}(x)$ each possess a large negative
contribution at $x=1$ (these are contained within the `plus prescription'
functions together with explicit delta functions in the definitions). This
contribution is included to take account of the fact that splittings of the
parton $i$ into other partons with lower momentum act to reduce the population
of partons with the original momentum. At a fundamental level, the
contributions at $x=1$ result from virtual gluon radiation diagrams.

On the other hand, the functions $P_{i \to jk}(x)$ do not contain such
contributions. This is to be expected as a virtual process is clearly not able
to achieve the $1 \to 2$ splitting $i \to jk$. At LO, the function $P_{i \to
jk}(x)$ is related to the `real splitting' 
part\footnote{The functions $P^R_{i \to j}(x)$ are obtained from the
functions $P_{i \to j}(x)$ by dropping the terms proportional to $\delta(1-x)$.
This includes removing plus prescription $+$ signs where they appear.} of the normal splitting
functions $P^{R}_{i \to j}(x)$ according to:
\begin{equation} \label{Splitting1}
P^{R}_{i \to j}(x) = \sum_k P_{i \to jk}(x)
\end{equation}
Equation~\eqref{Splitting1} is the simple statement that the probability of
splitting $i \to j + anything$ is equal to the sum of probabilities of
splitting $i \to j + k$, summed over all possibilities for $k$.

A further simplification to \eqref{Splitting1} is possible at LO. Due to the
fact that QCD only allows certain types of three particle vertices (i.e. triple
gluon vertices and `gluon emission from a quark' type vertices), the LO $P_{i
\to jk}(x)$ is only nonzero for a small number of $\{i,j,k\}$ combinations. In
fact, given $i$ and $j$, there exists at most one choice for $k$ which makes
$P_{i \to jk}(x)$ nonzero. We shall denote this special value of $k$ by
$\kappa(i,j)$. For example, $\kappa(i,j)$ is $g$ when $i=q_i,j=q_i$,
and $\bar{q_i}$ when $i=g, j=q_i$.

Given this fact, we note that \eqref{Splitting1} must contain at most only
one term on the right hand side, and we may write:
\begin{equation} \label{Splitting2}
P^{R}_{i \to j}(x) = P_{i \to j\kappa(i,j)}(x)
\end{equation}
In \eqref{Splitting2}, we have extended the definition of $\kappa(i,j)$ to
cases where there exists no choice for $k$ to make $P_{i \to jk}(x)$ nonzero.
In these cases, $\kappa(i,j)$ can be chosen to be any parton, as both the right
and left hand sides are zero for any choice.

Equation \eqref{Splitting2} effectively defines $P_{i \to jk}$ for all cases in
which it is nonzero. At LO then, we may construct the following definition for
$P_{i \to jk}$:
\begin{eqnarray} \label{Doubsplitdef}
P_{i \to jk}(x) = \begin{cases}
 P^R_{i \to j}(x) &\text{if $k$ = $\kappa(i,j)$}  \\
0 &\text{otherwise}
\end{cases}
\end{eqnarray}

It is interesting to consider the generalisation of the $P_{i \to jk}(x)$ functions
to NLO (and indeed higher orders). 
Here one encounters a problem, in that the function $P_{i \to jk}(x)$ only has one longitudinal
momentum argument because it is assumed that the parton $k$ must possess the
remaining longitudinal momentum originally carried by $i$ that was not given to
$j$, i.e. $1-x$. This is certainly true at leading order, where only two
partons can be produced in a single splitting, by conservation of momentum.
However, it is not true in general at NLO, where a single splitting can contain
two QCD vertices, and produce three partons. At NLO and above, the splitting
function $P_{i \to jk}$ should have two arguments, $x_1$ and $x_2$.

The expansion of the more general function $P_{i \to jk}(x_1,x_2)$ in terms of
powers of $\alpha_s$ would read as follows:
\begin{equation}
P_{i \to jk}(x_1,x_2) = \delta(1-x_1-x_2)P^{(0)}_{i \to jk}(x_1) +
\dfrac{\alpha_s}{2\pi}P^{(1)}_{i \to jk}(x_1,x_2) + \ldots
\end{equation}
The higher-order coefficients in this expansion cannot be obtained trivially
from the higher-order coefficients of the splitting function $P_{i\to j}(x)$ as
in the LO case. The general relation between the two for $x<1$ is:
\begin{equation}
P^{(n)}_{i \to j}(x_1) = \sum_k\int_0^{1-x_1} dx_2\; P^{(n)}_{i \to jk}(x_1,x_2)
\end{equation}
The unintegrated $P^{(n)}_{i \to jk}(x_1,x_2)$ must contain more information
than the $P^{(n)}_{i \to j}(x)$ for $n>0$, and hence cannot be obtained from
them.

A consequence of the fact that $P_{i \to jk}(x)$ with a single longitudinal
momentum argument $x$ is not the right function to use at NLO and above is that
the precise structure of the dDGLAP equation given in \eqref{dbDGLAP} can only
be applicable at LO.\footnote{This is not explicitly stated
in \cite{Kirschner:1979im, Shelest:1982dg}.} 
In what follows we restrict our analysis to the LO case, but will return to the generalisation of
 \eqref{dbDGLAP} to NLO in a future study.

One can interpret the terms on the right-hand side of \eqref{dbDGLAP} using the
parton branching picture.\footnote{We use similar arguments as are used
in Section~5.2 of \cite{Ellis:1996ws} to explain the terms on the right hand
side of the sDGLAP equation.} Consider the inclusive probability of finding a
pair of partons in the proton with flavours $j_1$ and $j_2$ and longitudinal
momentum fractions between $x_1$ and $x_1+\delta x_1$ and $x_2$ and $x_2+
\delta x_2$ respectively at scale $t$, $D_h^{j_1j_2}(x_1,x_2;t)\delta x_1
\delta x_2$. It is obvious that when $t$ is increased to $t+\Delta t$, two
types of process may contribute to the change in this quantity. Splittings from
higher-momentum partons giving rise to $j_1j_2$ pairs with the correct momentum
act to increase this quantity, whilst splittings within the $j_1j_2$ pairs to
give partons of lower momentum act to decrease the quantity.

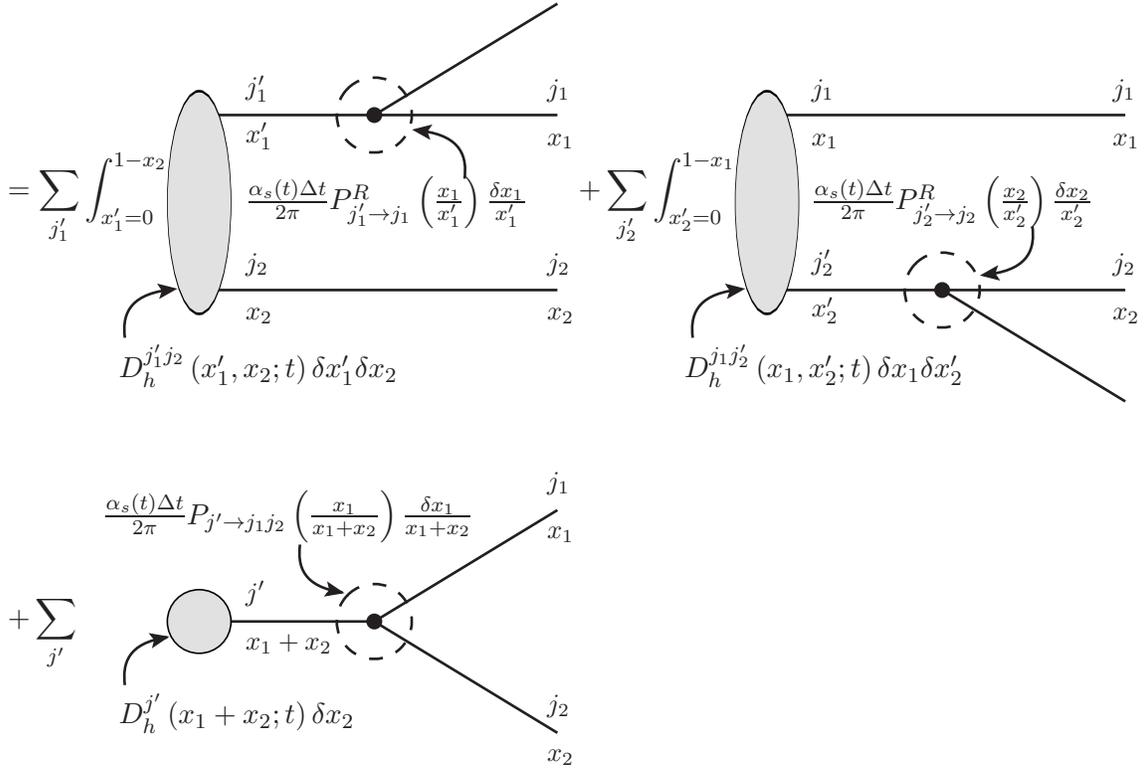
\begin{figure}
\centering
  \begin{picture}(468,315) (9,-3)
    \SetWidth{0.5}
    \SetColor{Black}
    \Text(6,291)[lb]{\normalsize{\Black{$\Delta_{+}\left[
D_{h}^{j_{1}j_{2}}\left(
x_{1},x_{2};t\right)\delta{x_{1}}\delta{x_{2}}\right]$}}}
    \Text(24,189)[lb]{\normalsize{\Black{$=\displaystyle\sum_{j'_{1}}\int_{x'_{1}=0}^{1-x_{2}}$}}}
    \SetWidth{1.0}
    \Line(96,237)(231,237)
    \Line(96,171)(231,171)
    \GOval(96,204)(42,12)(0){0.882}
    \Vertex(162,237){3}
    \Text(114,243)[lb]{\small{\Black{$j'_{1}$}}}
    \Text(114,225)[lb]{\small{\Black{$x'_{1}$}}}
    \Text(114,177)[lb]{\small{\Black{$j_{2}$}}}
    \Text(114,159)[lb]{\small{\Black{$x_{2}$}}}
    \Text(114,195)[lb]{\normalsize{\Black{$\frac{\alpha_{s}\left(t\right)\Delta{t}}{2\pi}P_{j'_{1}\to{j_{1}}}^{R}\left(\frac{x_{1}}{x'_{1}}\right)\frac{\delta{x_{1}}}{x'_{1}}$}}}
    \Arc[dash,dashsize=5](162,237)(14.142,135,495)
    \Text(228,177)[lb]{\small{\Black{$j_{2}$}}}
    \Text(228,159)[lb]{\small{\Black{$x_{2}$}}}
    \Text(228,243)[lb]{\small{\Black{$j_{1}$}}}
    \Text(228,225)[lb]{\small{\Black{$x_{1}$}}}
    \Bezier(180,231)(195,229)(197,221)(196,213)\Line[arrow,arrowpos=0.005,arrowlength=5,arrowwidth=2,arrowinset=0.2,flip](180,231)(180.224,230.97)%JaxoID:FBez[arrow,arrowpos=0.005,arrowlength=5,arrowwidth=2,arrowinset=0.2,flip]
    \Line(162,237)(231,279)
    \Text(66,135)[lb]{\normalsize{\Black{$D_{h}^{j'_{1}j_{2}}\left(x'_{1},x_{2};t\right)\delta{x'_{1}}\delta{x_{2}}$}}}
    \Bezier(84,171)(69,169)(67,161)(68,153)\Line[arrow,arrowpos=0.005,arrowlength=5,arrowwidth=2,arrowinset=0.2,flip](84,171)(83.776,170.97)%JaxoID:FBez[arrow,arrowpos=0.005,arrowlength=5,arrowwidth=2,arrowinset=0.2,flip]
    \Text(24,28)[lb]{\normalsize{\Black{$+\displaystyle\sum_{j'}$}}}
    \Line(102,46)(162,46)
    \Arc[dash,dashsize=5](162,46)(14.142,135,495)
    \Vertex(162,46){3}
    \Line(162,46)(231,4)
    \Line(162,46)(231,88)
    \Text(66,4)[lb]{\normalsize{\Black{$D_{h}^{j'}\left(x_{1}+x_{2};t\right)\delta{x_{2}}$}}}
    \Text(114,52)[lb]{\small{\Black{$j'$}}}
    \Text(114,34)[lb]{\small{\Black{$x_{1}+x_{2}$}}}
    \Text(60,76)[lb]{\normalsize{\Black{$\frac{\alpha_{s}\left(t\right)\Delta{t}}{2\pi}P_{j'\to{j_{1}j_{2}}}\left(\frac{x_{1}}{x_{1}+x_{2}}\right)\frac{\delta{x_{1}}}{x_{1}+x_{2}}$}}}
    \Text(228,94)[lb]{\small{\Black{$j_{1}$}}}
    \Text(228,76)[lb]{\small{\Black{$x_{1}$}}}
    \Text(228,10)[lb]{\small{\Black{$j_{2}$}}}
    \Text(228,-8)[lb]{\small{\Black{$x_{2}$}}}
    \SetWidth{0.6}
    \GOval(96,46)(12,12)(0){0.882}
    \SetWidth{1.0}
    \Bezier(150,57)(135,59)(133,67)(134,75)\Line[arrow,arrowpos=0.075,arrowlength=5,arrowwidth=2,arrowinset=0.2,flip](147.038,57.506)(146.84,57.549)%JaxoID:FBez[arrow,arrowpos=0.075,arrowlength=5,arrowwidth=2,arrowinset=0.2,flip]
    \Bezier(84,40)(69,38)(67,30)(68,22)\Line[arrow,arrowpos=0.075,arrowlength=5,arrowwidth=2,arrowinset=0.2,flip](81.038,39.494)(80.84,39.451)%JaxoID:FBez[arrow,arrowpos=0.075,arrowlength=5,arrowwidth=2,arrowinset=0.2,flip]
    \Text(240,189)[lb]{\normalsize{\Black{$+\displaystyle\sum_{j'_{2}}\int_{x'_{2}=0}^{1-x_{1}}$}}}
    \Line(310,171)(445,171)
    \Line(376,171)(445,129)
    \Line(310,237)(445,237)
    \Vertex(376,171){3}
    \Text(442,243)[lb]{\small{\Black{$j_{1}$}}}
    \Text(442,225)[lb]{\small{\Black{$x_{1}$}}}
    \Text(328,243)[lb]{\small{\Black{$j_{1}$}}}
    \Text(328,225)[lb]{\small{\Black{$x_{1}$}}}
    \Text(442,177)[lb]{\small{\Black{$j_{2}$}}}
    \Text(442,159)[lb]{\small{\Black{$x_{2}$}}}
    \Text(328,177)[lb]{\small{\Black{$j'_{2}$}}}
    \Text(328,159)[lb]{\small{\Black{$x'_{2}$}}}
    \Arc[dash,dashsize=5](376,171)(14.142,135,495)
    \Bezier(298,171)(283,169)(281,161)(282,153)\Line[arrow,arrowpos=0.005,arrowlength=5,arrowwidth=2,arrowinset=0.2,flip](298,171)(297.776,170.97)%JaxoID:FBez[arrow,arrowpos=0.005,arrowlength=5,arrowwidth=2,arrowinset=0.2,flip]
    \GOval(310,204)(42,12)(0){0.882}
    \Text(280,135)[lb]{\normalsize{\Black{$D_{h}^{j_{1}j'_{2}}\left(x_{1},x'_{2};t\right)\delta{x_{1}}\delta{x'_{2}}$}}}
    \Text(328,195)[lb]{\normalsize{\Black{$\frac{\alpha_{s}\left(t\right)\Delta{t}}{2\pi}P_{j'_{2}\to{j_{2}}}^{R}\left(\frac{x_{2}}{x'_{2}}\right)\frac{\delta{x_{2}}}{x'_{2}}$}}}
    \Bezier(394,177)(409,179)(411,187)(410,195)\Line[arrow,arrowpos=0.005,arrowlength=5,arrowwidth=2,arrowinset=0.2,flip](394,177)(394.224,177.03)%JaxoID:FBez[arrow,arrowpos=0.005,arrowlength=5,arrowwidth=2,arrowinset=0.2,flip]
  \end{picture}
\caption{\label{fig:insplittings}Splitting processes that increase the
population of $j_1j_2$ parton pairs with momenta in the ranges $x_1 \to
x_1+\delta x_1$, $x_2 \to x_2+\delta x_2$. $P^R_{i \to j}(x)$ is the `real
splitting' part of the splitting function $P_{i \to j}(x)$ -- i.e. the splitting
function minus the terms proportional to $\delta(1-x)$.}
\end{figure}

At leading order in $\alpha_s$ (which, as we have established, is the order
under which equation \eqref{dbDGLAP} was derived), there are three types of
splitting process that give rise to a pair of partons $j_1j_2$ with momenta in
the ranges $x_1 \to x_1+\delta x_1$, $x_2 \to x_2+\delta x_2$. The three are
drawn schematically in Fig.~\ref{fig:insplittings}.

\begin{figure}
 \begin{picture}(437,184) (9,-3)
    \SetWidth{0.5}
    \SetColor{Black}
    \Text(6,160)[lb]{\normalsize{\Black{$\Delta_{-}\left[
D_{h}^{j_{1}j_{2}}\left(
x_{1},x_{2};t\right)\delta{x_{1}}\delta{x_{2}}\right]$}}}
    \Text(42,70)[lb]{\normalsize{\Black{$=$}}}
    \SetWidth{1.0}
    \Line(96,106)(231,106)
    \Line(96,40)(231,40)
    \GOval(96,73)(42,12)(0){0.882}
    \Vertex(162,106){3}
    \Text(114,112)[lb]{\small{\Black{$j_{1}$}}}
    \Text(114,94)[lb]{\small{\Black{$x_{1}$}}}
    \Text(114,46)[lb]{\small{\Black{$j_{2}$}}}
    \Text(114,28)[lb]{\small{\Black{$x_{2}$}}}
    \Text(156,64)[lb]{\normalsize{\Black{$\frac{\alpha_{s}\left(t\right)\Delta{t}}{2\pi}P_{j_{1}\to{j_{1}}}^{V}$}}}
    \Arc[dash,dashsize=5](162,106)(14.142,135,495)
    \Bezier(180,100)(195,98)(197,90)(196,82)\Line[arrow,arrowpos=0.005,arrowlength=5,arrowwidth=2,arrowinset=0.2,flip](180,100)(180.224,99.97)%JaxoID:FBez[arrow,arrowpos=0.005,arrowlength=5,arrowwidth=2,arrowinset=0.2,flip]
    \Line(162,106)(231,148)
    \Text(66,4)[lb]{\normalsize{\Black{$D_{h}^{j_{1}j_{2}}\left(x_{1},x_{2};t\right)\delta{x_{1}}\delta{x_{2}}$}}}
    \Bezier(84,40)(69,38)(67,30)(68,22)\Line[arrow,arrowpos=0.005,arrowlength=5,arrowwidth=2,arrowinset=0.2,flip](84,40)(83.776,39.97)%JaxoID:FBez[arrow,arrowpos=0.005,arrowlength=5,arrowwidth=2,arrowinset=0.2,flip]
    \Text(256,70)[lb]{\normalsize{\Black{$+$}}}
    \Line(310,40)(445,40)
    \Line(376,40)(445,-2)
    \Line(310,106)(445,106)
    \Vertex(376,40){3}
    \Arc[dash,dashsize=5](376,40)(14.142,135,495)
    \Bezier(298,40)(283,38)(281,30)(282,22)\Line[arrow,arrowpos=0.005,arrowlength=5,arrowwidth=2,arrowinset=0.2,flip](298,40)(297.776,39.97)%JaxoID:FBez[arrow,arrowpos=0.005,arrowlength=5,arrowwidth=2,arrowinset=0.2,flip]
    \GOval(310,73)(42,12)(0){0.882}
    \Text(280,4)[lb]{\normalsize{\Black{$D_{h}^{j_{1}j_{2}}\left(x_{1},x_{2};t\right)\delta{x_{1}}\delta{x_{2}}$}}}
    \Text(370,70)[lb]{\normalsize{\Black{$\frac{\alpha_{s}\left(t\right)\Delta{t}}{2\pi}P_{j_{2}\to{j_{2}}}^{V}$}}}
    \Bezier(394,46)(409,48)(411,56)(410,64)\Line[arrow,arrowpos=0.005,arrowlength=5,arrowwidth=2,arrowinset=0.2,flip](394,46)(394.224,46.03)%JaxoID:FBez[arrow,arrowpos=0.005,arrowlength=5,arrowwidth=2,arrowinset=0.2,flip]
    \Text(329,112)[lb]{\small{\Black{$j_{1}$}}}
    \Text(329,94)[lb]{\small{\Black{$x_{1}$}}}
    \Text(329,46)[lb]{\small{\Black{$j_{2}$}}}
    \Text(329,28)[lb]{\small{\Black{$x_{2}$}}}
  \end{picture}
\caption{\label{fig:outsplittings}Splitting processes that decrease the
population of $j_1j_2$ parton pairs with momenta in the ranges $x_1 \to
x_1+\delta x_1$, $x_2 \to x_2+\delta x_2$. $P^V_{j \to j}$ is equal to the sum
of the coefficients of the $\delta(1-x)$ terms in the splitting function $P_{j
\to j}(x)$ (including $\delta(1-x)$ terms contained within plus prescription
functions).}
\end{figure}

In the first, we start with a pair of partons $j_1'j_2$ with momenta in the
ranges $x'_1 \to x'_1+\delta x'_1$, $x_2 \to x_2+\delta x_2$. The quantity
$x_1'$ must satisfy $x_1<x_1'<1-x_2$ -- i.e. be large enough that $j_1'$ can
split to give $j_1$, and be small enough that the initial pair of partons is
not carrying more momentum than the proton they are in. The parton $j_1'$ then
splits to give as one of the products a $j_1$ with momentum in the range $x_1
\to x_1+\delta x_1$. The second process is very similar but involves a
splitting in the second parton. The third involves a single parton $j'$ with
just the right momentum $x_1+x_2 \to x_1+x_2 + \delta x_2$ splitting to give as
its two daughters the pair $j_1j_2$ with momenta in the appropriate ranges.

The leading order splitting processes reducing the population of $j_1j_2$
partons with momenta in the given ranges are given in 
Fig.~\ref{fig:outsplittings}. There are two processes -- in the first, the $j_1$
parton splits to give lower momentum partons, whilst in the second the $j_2$
splits.

The correspondence between the diagrams of Figs.~\ref{fig:insplittings} and
\ref{fig:outsplittings} representing the splitting processes and the terms on
the RHS of \eqref{dbDGLAP} is fairly clear. Suitable labels have been added to
the figures to bring out this correspondence. It is important to note that the
first two sets of terms on the RHS of \eqref{dbDGLAP} cover four diagrams -- the
first two of both Fig.~\ref{fig:insplittings} and Fig.~\ref{fig:outsplittings}. 
The `real splitting' parts of the terms correspond to
the diagrams in Fig.~\ref{fig:insplittings}, whilst the `virtual correction'
parts correspond to the diagrams in Fig.~\ref{fig:outsplittings}. We also
note that the last set of terms on the RHS of \eqref{dbDGLAP} contains sPDFs
because it corresponds to the diagram in which a {\em single} parton splits to
give the pair $j_1j_2$. There is no integral in these terms because of the
property of LO QCD that a single splitting can only give rise to two partons.
Thus the single parton that splits is essentially restricted to have momentum
exactly equal to $x_1+x_2$. We shall hereafter refer to the last set of terms on
the RHS of \eqref{dbDGLAP} as the `sPDF feed' terms, for obvious reasons.

A solution to \eqref{dbDGLAP} in terms of sPDFs is obtained in
\cite{Kirschner:1979im, Shelest:1982dg}, and presented for the first time in
$x$-space in \cite{Snigirev:2003cq}. Let us introduce the `natural' evolution
variable $\tau$ defined in terms of $t$ according to:
\begin{eqnarray}
\tau &=& \int_{t_0}^t dt' \dfrac{\alpha_s(t')}{2\pi} \\ \nonumber
     &=& \dfrac{1}{2\pi b} \ln \left[
\dfrac{t-\ln(\Lambda_{QCD}^2)}{t_0-\ln(\Lambda_{QCD}^2)} \right]
\hspace{5pt} \text{at LO}
\end{eqnarray}
In terms of the variable $\tau$, the solution to the dDGLAP equation reads:
\begin{eqnarray} \label{dbDGLAPsoln}
D^{{j_1}{j_2}}_h(x_1,x_2;\tau) &=& D_{h(corr)}^{j_1j_2}(x_1,x_2;\tau)
\\ \nonumber
&+&\sum_{j'_1j'_2}\int_{x_1}^{1-x_2}\dfrac{dz_1}{z_1}\int_{x_2}^{1-z_1}
\dfrac{dz_2}{z_2}D_h^{j'_1j'_2}(z_1,z_2;\tau =0)
 \\ \nonumber
&&\times D_{j_1'}^{j_1}\left(\frac{x_1}{z_1};\tau\right)
D_{j_2'}^{j_2}\left(\dfrac{x_2}{z_2};\tau\right)
\end{eqnarray}

\noindent where:
\begin{eqnarray} \label{defDcorr}
D_{h(corr)}^{j_1j_2}(x_1,x_2;\tau) =
\sum_{j'j'_1j'_2}\int_{0}^{\tau}d\tau '
\int_{x_1}^{1-x_2}\frac{dz_1}{z_1}\int_{x_2}^{1-z_1}\frac{dz_2}{z_2}
D_h^{j'}(z_1+z_2;\tau ') \\ \nonumber
\times \frac{1}{z_1+z_2}
P_{j' \to j'_1j'_2}\left(\frac{z_1}{z_1+z_2}\right)
D_{j_1'}^{j_1}\left(\frac{x_1}{z_1};\tau-\tau '\right)
D_{j_2'}^{j_2}\left(\frac{x_2}{z_2};\tau-\tau '\right)
\end{eqnarray}

The Green's functions $D_{i}^{j}\left(x;\tau\right)$ are defined such that they
satisfy the initial conditions
$D_{i}^{j}\left(x;\tau=0\right)=\delta_{ij}\delta(1-x)$ and change with $\tau$
according to the sDGLAP equation:
\begin{equation}
\dfrac{dD_{i}^{j}(x;\tau)}{d\tau}=\sum_{j'}\int_x^1\dfrac{dz}{z}D_i^{j'}(z;\tau)P_{j'\to
j}(x/z)
\end{equation}
In effect, the function $D_{i}^{j}\left(x;\tau\right)$ gives the inclusive
probability that one finds a parton $j$ with longitudinal momentum fraction $x$
at scale $\tau$ inside a dressed object that looks like a pure $i$ parton at
the scale $\tau = 0$.

\begin{figure}
  \begin{picture}(452,406) (9,-3)
    \SetWidth{0.5}
    \SetColor{Black}
    \Text(6,382)[lb]{\normalsize{\Black{$D_{h}^{j_{1}j_{2}}\left(x_{1},x_{2};\tau\right)\delta{x_{1}}\delta{x_{2}}$}}}
    \Text(42,46)[lb]{\normalsize{\Black{$+\displaystyle\sum_{j'_{1},j'_{2}}\int_{z_{1}=x_{1}}^{1-x_{2}}\int_{z_{2}=x_{2}}^{1-z_{1}}$}}}
    \Text(41,280)[lb]{\normalsize{\Black{$=\displaystyle\sum_{j',j'_{1},j'_{2}}\int_{\tau'=0}^{\tau}\int_{z_{1}=x_{1}}^{1-x_{2}}\int_{z_{2}=x_{2}}^{1-z_{1}}$}}}
    \SetWidth{1.0}
    \Line(287,292)(356,334)
    \Line(354,334)(426,377)
    \Line(287,292)(356,250)
    \Line(354,250)(426,207)
    \SetWidth{0.6}
    \GOval(354,250)(12,12)(0){0.882}
    \SetWidth{1.0}
    \Line(227,292)(287,292)
    \Vertex(287,292){3}
    \Text(191,250)[lb]{\normalsize{\Black{$D_{h}^{j'}\left(z_{1}+z_{2};\tau'\right)\delta{z_{2}}$}}}
    \Text(239,298)[lb]{\small{\Black{$j'$}}}
    \Text(239,280)[lb]{\small{\Black{$z_{1}+z_{2}$}}}
    \Text(162,316)[lb]{\normalsize{\Black{$\Delta\tau{P_{j'\to{j'_{1}j'_{2}}}\left(\frac{z_{1}}{z_{1}+z_{2}}\right)\frac{\delta{z_{1}}}{z_{1}+z_{2}}}$}}}
    \SetWidth{0.6}
    \GOval(221,292)(12,12)(0){0.882}
    \SetWidth{1.0}
    \Bezier(282,298)(267,300)(265,308)(266,316)\Line[arrow,arrowpos=0.075,arrowlength=5,arrowwidth=2,arrowinset=0.2,flip](279.038,298.506)(278.84,298.549)%JaxoID:FBez[arrow,arrowpos=0.075,arrowlength=5,arrowwidth=2,arrowinset=0.2,flip]
    \Bezier(209,286)(194,284)(192,276)(193,268)\Line[arrow,arrowpos=0.075,arrowlength=5,arrowwidth=2,arrowinset=0.2,flip](206.038,285.494)(205.84,285.451)%JaxoID:FBez[arrow,arrowpos=0.075,arrowlength=5,arrowwidth=2,arrowinset=0.2,flip]
    \SetWidth{0.6}
    \GOval(354,334)(12,12)(0){0.882}
    \Text(300,364)[lb]{\normalsize{\Black{$D_{j'_{1}}^{j_{1}}\left(\frac{x_{1}}{z_{1}};\tau-\tau'\right)\frac{\delta{x_{1}}}{z_{1}}$}}}
    \Text(300,202)[lb]{\normalsize{\Black{$D_{j'_{2}}^{j_{2}}\left(\frac{x_{2}}{z_{2}};\tau-\tau'\right)\frac{\delta{x_{2}}}{z_{2}}$}}}
    \SetWidth{1.0}
    \Line[dash,dashsize=6](426,190)(426,394)
    \Line[dash,dashsize=6](288,190)(288,394)
    \Bezier(342,340)(327,342)(325,350)(326,358)\Line[arrow,arrowpos=0.075,arrowlength=5,arrowwidth=2,arrowinset=0.2,flip](339.038,340.506)(338.84,340.549)%JaxoID:FBez[arrow,arrowpos=0.075,arrowlength=5,arrowwidth=2,arrowinset=0.2,flip]
    \Bezier(342,244)(327,242)(325,234)(326,226)\Line[arrow,arrowpos=0.075,arrowlength=5,arrowwidth=2,arrowinset=0.2,flip](339.038,243.494)(338.84,243.451)%JaxoID:FBez[arrow,arrowpos=0.075,arrowlength=5,arrowwidth=2,arrowinset=0.2,flip]
    \Text(288,172)[lb]{\normalsize{\Black{$\tau'$}}}
    \Text(426,172)[lb]{\normalsize{\Black{$\tau$}}}
    \Text(318,322)[lb]{\small{\Black{$j'_{1}$}}}
    \Text(318,304)[lb]{\small{\Black{$z_{1}$}}}
    \Text(318,280)[lb]{\small{\Black{$j'_{2}$}}}
    \Text(318,262)[lb]{\small{\Black{$z_{2}$}}}
    \Text(414,376)[lb]{\small{\Black{$j_{1}$}}}
    \Text(414,358)[lb]{\small{\Black{$x_{1}$}}}
    \Text(414,220)[lb]{\small{\Black{$j_{2}$}}}
    \Text(414,202)[lb]{\small{\Black{$x_{2}$}}}
    \Line(166,28)(301,28)
    \Line(166,94)(301,94)
    \Bezier(154,28)(139,26)(137,18)(138,10)\Line[arrow,arrowpos=0.005,arrowlength=5,arrowwidth=2,arrowinset=0.2,flip](154,28)(153.776,27.97)%JaxoID:FBez[arrow,arrowpos=0.005,arrowlength=5,arrowwidth=2,arrowinset=0.2,flip]
    \GOval(166,61)(42,12)(0){0.882}
    \Text(138,-8)[lb]{\normalsize{\Black{$D_{h}^{j'_{1}j'_{2}}\left(z_{1},z_{2};0\right)\delta{z_{1}}\delta{z_{2}}$}}}
    \SetWidth{0.6}
    \GOval(234,28)(12,12)(0){0.882}
    \GOval(234,94)(12,12)(0){0.882}
    \Text(204,52)[lb]{\normalsize{\Black{$D_{j'_{2}}^{j_{2}}\left(\frac{x_{2}}{z_{2}};\tau\right)\frac{\delta{x_{2}}}{z_{2}}$}}}
    \Text(204,118)[lb]{\normalsize{\Black{$D_{j'_{1}}^{j_{1}}\left(\frac{x_{1}}{z_{1}};\tau\right)\frac{\delta{x_{1}}}{z_{1}}$}}}
    \SetWidth{1.0}
    \Bezier(222,100)(207,102)(205,110)(206,118)\Line[arrow,arrowpos=0.075,arrowlength=5,arrowwidth=2,arrowinset=0.2,flip](219.038,100.506)(218.84,100.549)%JaxoID:FBez[arrow,arrowpos=0.075,arrowlength=5,arrowwidth=2,arrowinset=0.2,flip]
    \Bezier(222,34)(207,36)(205,44)(206,52)\Line[arrow,arrowpos=0.075,arrowlength=5,arrowwidth=2,arrowinset=0.2,flip](219.038,34.506)(218.84,34.549)%JaxoID:FBez[arrow,arrowpos=0.075,arrowlength=5,arrowwidth=2,arrowinset=0.2,flip]
    \Text(180,100)[lb]{\small{\Black{$j'_{1}$}}}
    \Text(180,82)[lb]{\small{\Black{$z_{1}$}}}
    \Text(180,34)[lb]{\small{\Black{$j'_{2}$}}}
    \Text(180,16)[lb]{\small{\Black{$z_{2}$}}}
    \Text(294,100)[lb]{\small{\Black{$j_{1}$}}}
    \Text(294,82)[lb]{\small{\Black{$x_{1}$}}}
    \Text(294,34)[lb]{\small{\Black{$j_{2}$}}}
    \Text(294,16)[lb]{\small{\Black{$x_{2}$}}}
  \end{picture}
\caption{\label{fig:dbDGLAPsoln} A schematic representation of the solution of
the dDGLAP equation \eqref{dbDGLAPsoln} in terms of the parton branching
picture.}
\end{figure}

A pictorial representation of the solution \eqref{dbDGLAPsoln} in terms of
parton branching is given in Fig.~\ref{fig:dbDGLAPsoln}. One observes the
need to specify some initial conditions $D_h^{j'_1j'_2}(x_1,x_2;\tau =0)$ to
obtain the distributions at higher scale, which is a direct reflection of the
fact that the dDGLAP equation can only predict changes in the distributions
with $\tau$.

The depiction of dDGLAP evolution as in Fig.~\ref{fig:dbDGLAPsoln} leads us to 
make a suggestion as to how one might calculate the double distributions for which 
the two scales are not equal, $D_h^{ij}(x_1,x_2;\tau_1,\tau_2)$. The arguments $\tau_1$
and $\tau_2$ in this distribution correspond to the factorisation scales for parton
$i$ and $j$ respectively. Consider the analogous figure to Fig.~\ref{fig:dbDGLAPsoln} 
for these distributions. It seems
likely that this figure would be the same, except with $\tau_1$ replacing $\tau$
on the `upper legs' of the diagrams, $\tau_2$ replacing $\tau$ on the `lower legs' 
of the diagrams, and the upper limit of the $\tau'$ integration replaced by 
$\rm{min}(\tau_1,\tau_2)$. If this ansatz is correct, the double distributions
$D_h^{ij}(x_1,x_2;\tau_1,\tau_2)$ with (say) $\tau_1 < \tau_2$ should be calculated
by taking the dPDFs with $\tau = \tau_1$, and then performing sDGLAP evolution 
at each $x_1$ from $\tau_1$ to $\tau_2$ in the $x_2$ variable. The upper limit in the
sDGLAP evolution at given $x_1$ should be $1-x_1$.

\newpage
\hspace{1mm}
\newpage

\section{The Double Parton Sum Rules and the Initial Distributions}
\label{sec:sumrules}

\subsection{The Double Parton Sum Rules} \label{sec:sumrulesgen}

It is well known that the sPDFs satisfy two types of sum rules which represent
the fact that both momentum and valence quark number should be conserved under
evolution. One might wonder whether corresponding rules exist for the dPDFs. We 
have managed to prove that the following equalities are preserved by dDGLAP evolution,
provided they hold at the starting scale $t_0$:
\\

\noindent {\em Momentum Sum Rule}: \\
Let $M$ be the momentum fraction carried by the proton ($=1$). Then:
\begin{equation} \label{mtmsum}
\sum_{j_1}\int_0^{1-x_2} dx_1 x_1 D^{{j_1}{j_2}}_h(x_1,x_2;t) =
(M-x_2)D_h^{j_2}(x_2;t)
\end{equation}

\noindent {\em Number Sum Rule}: \\
Let $j_{1v} \equiv j_1-\overline{j}_1$ ($j_1\ne g$), and $N_{j_{1v}}$ be the
number of `valence' $j_1$ quarks in the proton. Then:
\begin{eqnarray} \label{numsum}
\int_0^{1-x_2}dx_1D_h^{j_{1v}j_2}(x_1,x_2;t)=\begin{cases}
N_{j_{1v}} D_h^{j_2}(x_2;t) & \text{when $j_2 \ne j_1$ or $\overline{j}_1$} \\
(N_{j_{1v}}-1) D_h^{j_2}(x_2;t) & \text{when $j_2 = j_1$} \\
(N_{j_{1v}}+1) D_h^{j_2}(x_2;t) & \text{when $j_2 = \overline{j}_1$}
\end{cases}
\end{eqnarray}
\\

The only nontrivial inputs to this proof are the following relations, which
must be obeyed by the splitting functions in order that the number and momentum
integrals are conserved for the sPDFs:
\begin{equation}
\sum_{j}\int_0^{1}dx_1 x_1 P_{j' \to j}\left(x_1\right) = 0; 
\quad \int_0^{1}dx_1 P_{j' \to j_{v}}\left(x_1\right) = 0
\end{equation}
We do not present the proof here, since it is straightforward and rather lengthy.
A crucial point we must emphasise however is that the precise structure of the 
sum rules given above is required in order that they should be 
preserved by dDGLAP evolution. In particular, the prefactors in front of the 
sPDF quantities on the right hand sides of \eqref{mtmsum} and \eqref{numsum}
 must have the values given. 

By appropriately combining equations \eqref{mtmsum} and \eqref{numsum} with the 
sPDF momentum and number sum rules, one can construct integrals over both 
arguments of the dPDFs which give conserved quantities such as $M$ or $N_{j_v}$
(or products of these quantities). Examples of such integrals are given below:

\begin{eqnarray}
\sum_{j_1j_2}\int_0^1 dx_2 \int_0^{1-x_2} dx_1 
\dfrac{x_1 x_2}{M-x_2} D^{{j_1}{j_2}}_h(x_1,x_2;t) =
M = 1 \\
\int_0^1 dx_2 \int_0^{1-x_2} dx_1 \dfrac{D_h^{j_{1v}j_1}(x_1,x_2;t)}{N_{j_{1v}}-1} 
- \dfrac{D_h^{j_{1v}\bar{j}_1}(x_1,x_2;t)}{N_{j_{1v}}+1}
= N_{j_{1v}}
\end{eqnarray}

These relations are preserved under dDGLAP evolution. By contrast,
integrals such as $\sum_{j_1j_2}\int_0^1 dx_2 \int_0^{1-x_2} dx_1 
x_1 x_2 D^{{j_1}{j_2}}_h(x_1,x_2;t)$ and $\int_0^1 dx_2 \int_0^{1-x_2} dx_1 
D_h^{j_{1v}j_{1v}}(x_1,x_2;t)$, which one might naively think should give conserved
momenta or valence quark numbers, are not conserved by dDGLAP evolution and so
do not correspond to such physical quantities.

An appealing interpretation of \eqref{mtmsum} and \eqref{numsum} exists in terms
of probability theory (although such a picture has in no way been used to obtain
these relations). The dPDF sum rules are analagous to the result in probability
theory that for two continuous random variables $X$ and $Y$,  the
probability density functions relating to $X$ and $Y$ must satisfy 
\footnote{One should bear 
in mind that the correspondence between \eqref{condprob} and \eqref{mtmsum}/\eqref{numsum} is 
not completely straightforward, as the parton density functions are not really simple
probabilities. Rather, they may be better interpreted as number distributions. This
results in, for example, the sPDFs being normalised to the number of partons of the 
given type in the proton rather than $1$. Of course an analogous relation to 
\eqref{condprob} exists for such distributions.}:
\begin{eqnarray} \label{condprob}
\int dx x^a f(X=x \cap Y=y) = E(X^a \mid Y=y) f(Y=y)
\end{eqnarray}

The integral is performed over all values that $X$ can take
given that $Y=y$, and $E(X^a \mid Y=y)$ is the
expectation value of $X^a$ given that $Y$ has value $y$. All of the prefactors
on the right hand sides of Eqns.~\eqref{mtmsum} and \eqref{numsum} are 
essentially conditional expectations as in \eqref{condprob}. The $(1-x_2)$ factor
on the right hand side of \eqref{mtmsum} is the conditional expectation value for the
 momentum of all of the other partons in the proton given that one has found a parton of 
longitudinal momentum fraction $x_2$. The $(N_{j_{1v}}-1)$ factor for the $j_2=j_1$ 
case of \eqref{numsum} is the conditional expectation for the number of $j_1$
partons minus the number of $\bar{j}_1$ partons elsewhere in the proton, given that
one has found a parton of flavour $j_1$. The prefactors for the other number
sum rule cases may be interpreted as conditional expectation values using
similar logic.

The fact that the forms of the number and momentum sum rules can be justified
using general arguments strongly suggests that these rules should hold to all
orders in perturbation theory, just as the sPDF number and momentum sum rules
do. We have been restricted to an LO proof that they hold at all scales if they
hold at the starting scale by the fact that we only have the LO dDGLAP
equation.

Although we have not derived the momentum and number sum rules from first principles,
they appear to satisfy a number of non-trivial consistency checks. For example, one
might worry that there might not be a set of dPDFs that satisfy the full
set of rules. A potential source of tension between the different rules is the
integral:
\begin{equation}
\sum_{j_2}\int_0^1dx_1\int_0^{1-x_1}dx_2x_2D^{j_{1v}j_2}(x_1,x_2;t)
\end{equation}

One can evaluate this integral using \eqref{mtmsum} or \eqref{numsum}, in
combination with appropriate sPDF sum rules. If different results were produced
depending on which dPDF sum rule was used, this would indicate an
inconsistency. However, one obtains the same result, $N_{j_{1v}}-f_{j_{1v}}$
(where $f_{j_{1v}}$ is the momentum fraction carried by valence $j_1$ partons),
with either approach. This lends further credibility to the sum rules
\eqref{mtmsum} and \eqref{numsum}.

It is notable that the complete set of dPDF sum rules, \eqref{mtmsum} and \eqref{numsum}, do not appear anywhere in the extant literature, although similar sum rules have been derived for the two-particle fragmentation functions in \cite{Konishi:1979cb}. An early paper on the subject, \cite{Goebel:1979mi} (see also \cite{Halzen:1986ue}), introduces some `constraints' resembling the number sum rules, which are used as an aid in constructing some simple model dPDFs. However, the constraints are only imposed for two specific dPDF cases, and the paper does not make any explicit statement about the general form of the number sum rule. In particular, they do not describe the subtleties of the number sum rule with regard to the different possible proportionality constants on the right hand side of \eqref{numsum}.

In some sense, the dPDF sum rules are more restrictive than their sPDF
counterparts. The sPDF sum rules state that the quantities $M \equiv
\sum_i\int_0^1 dx x D_h^i(x;t)$ and $N_{i_v} \equiv \int_0^1 dx D_h^{i_v}(x;t)$
are conserved under evolution whatever their initial values, and we make the
physical choices $M=1, N_{u_v} = 2, N_{d_v} = 1$ for the proton. On the other
hand, Eqns.~\eqref{mtmsum} and \eqref{numsum} are only preserved under
evolution if they hold at the starting scale. This is linked to the fact that
one initially has the freedom in the sum rules to specify the momentum/parton
composition of the hadron $M$ and $N_{i_v}$ (although $M \ne 1$ is not very
physical). However, once these have been specified in the sPDF sector, the
structure of the multiparton sum rules is effectively fixed.

The restrictive nature of the dPDF sum rules can be used to place nontrivial
constraints on the input distributions that are physically allowable in the dDGLAP
equation. If we believe that the dPDF sum rules should hold at the starting
scale, then we can use the constraints provided by the rules to improve on the factorised
inputs previously used at the starting scale $Q_0^2 \sim 1$~GeV$^2$. This is
discussed in the next section.

\subsection{Use of the Double Parton Sum Rules to improve the Input
Distributions}\label{sec:sumrulesinput}

As was mentioned in Section~\ref{sec:intro}, it is a common assumption that the
input double distributions should be equal to the product of the relevant sPDFs
at low $x_1$ and $x_2$. The logic behind this is that there exist large
populations of partons of all active flavour types and $x$ values at low $x$.
Given these large populations, we would expect the extraction of a parton with
a given flavour type $j_1$ and small longitudinal momentum $x_1$ not to have a
strong effect on the probability of finding another parton of flavour $j_2$
(where $j_2$ can be equal to $j_1$) and small longitudinal momentum $x_2$. This
leads to a joint probability distribution which can be expressed as a product
of single distributions at low $x_1$, $x_2$.

This factorisation  assumption appears to 
be backed up by the available CDF and D0 data. Consequently, we would like our
improved input dPDFs to maintain a factorised form for low $x_1,x_2$, whilst
now obeying the sum rules \eqref{mtmsum} and \eqref{numsum}. The first question
to be addressed in this section is whether this is in fact possible for all the dPDFs,
i.e. whether the sum rules are compatible with factorisation at low $x_1,x_2$ in all
cases.

\begin{table}[t]
\centering
\begin{tabular}{|l|l|}
\hline
dPDF Type & Relevant Sum Rules \\
\hline
Valence-Valence & Number (involved in two rules) \\
Valence-Sum     & Number + Momentum     \\
Valence-Tensor  & Number                \\
Tensor-Tensor   & None                  \\
Tensor-Sum      & Momentum              \\
Sum-Sum         & Momentum (involved in two rules) \\
\hline
\end{tabular}
\caption{\label{tab:dPDFtypes} The different dPDF classes under the `double
evolution' representation of the dPDFs, and the types of sum rules each is
engaged in.}
\end{table}

To help answer this question, we introduce the `double
evolution' representation for the dPDFs. In this representation, the well-known
\{singlet,gluon,valence,tensor\}  /$\{\Sigma,g,V_i,T_i\}$ combinations (defined
in, for example, Chapter 4 of \cite{Ellis:1996ws}) are used as the flavour
basis for both parton indices in the dPDF. The relationship between this basis and the `double human' basis in which both parton indices $i,j$ are one of $g, u, \bar{u}$ etc. can be clarified using an example:
\begin{eqnarray} \label{ev2hum}
D_h^{T_3 u_v} &=& D_h^{(u+\bar{u}-d-\bar{d})(u-\bar{u})} \\ \nonumber
              &=& D_h^{uu} + D_h^{\bar{u}u} - D_h^{du} - D_h^{\bar{d}u} 
                - D_h^{u\bar{u}} - D_h^{\bar{u}\bar{u}} + D_h^{d\bar{u}} + D_h^{\bar{d}\bar{u}}
\end{eqnarray}

The longitudinal momentum arguments of each term in this equation are the same. The use of the `double evolution' representation has
the advantage that it splits the dPDFs into six sets, each of which must
satisfy different combinations of the sum rules. We refer to the singlet
and gluon combinations as the `sum' combinations (as they describe the sum of
quark and gluon contributions respectively). Since $\sum j = \Sigma+g$, any
dPDF with a `sum' flavour index will be involved in a momentum sum rule, whilst
any dPDF with a `valence' flavour index will be involved in a number sum rule.
Those dPDFs where each of the indices are one out of the `sum' and `valence'
combinations will be involved in two sum rules. 

The six sets of dPDFs along
with the combinations of sum rules each is involved in are given in 
Table~\ref{tab:dPDFtypes}. We do not write out the explicit forms of the sum 
rules under the double evolution basis in this table. To obtain each rule, one must first construct the appropriate integral (i.e.~$\int dx_1 x_1 [D_h^{\Sigma k}(x_1,x_2) 
+ D_h^{g k}(x_1,x_2)]$ for a momentum sum rule or $\int dx_1 x_1 D_h^{i_v k}(x_1,x_2)$
for a number sum rule, where $k$ can be any double evolution basis index). 
The sum rule is then obtained by expanding each dPDF in the integral in terms of human basis dPDFs (as in \eqref{ev2hum}), followed by the use of equations \eqref{mtmsum} and \eqref{numsum}. We illustrate this procedure for the case of the $u_v T_3$ number sum rule:
\begin{align}
\int_0^{1-x_1}dx_2 D_h^{T_3 u_v}(x_1,x_2) =& \int_0^{1-x_1}dx_2  \Big[  D_h^{uu_v}(x_1,x_2) + D_h^{\bar{u}u_v}(x_1,x_2) \\ \nonumber
& \hspace{15mm} - D_h^{du_v}(x_1,x_2) - D_h^{\bar{d}u_v}(x_1,x_2) \Big]  \\ \nonumber
=& (N_{u_v}-1)D_h^{u}(x_1) + (N_{u_v}+1)D_h^{\bar{u}}(x_1) \\ \nonumber
&- N_{u_v}D_h^{d}(x_1) - N_{u_v}D_h^{\bar{d}}(x_1) \\ \nonumber
=& N_{u_v}D_h^{T_3}(x_1)-D_h^{u_v}(x_1)
\end{align}

If one investigates each class of dPDF and their respective sum rules, then one
finds that in most cases one is allowed dPDFs which satisfy the sum rules and
are approximately equal to the product of single distributions at low $x_1$ and
$x_2$. There is however a type of dPDF for which these two requirements cannot
be simultaneously satisfied -- the dPDF with two of the same valence
combinations as its flavour indices (e.g. $D_h^{u_vu_v}$). 

The number sum rule that this type of dPDF must satisfy reads:
\begin{equation} \label{valvalsum}
\int_0^{1-x_2}dx_1D_h^{j_vj_v}(x_1,x_2;t_0) = 
N_{j_v}D_h^{j_v}(x_2;t_0)-D_h^{j+\bar{j}}(x_2;t_0)
\end{equation}
Consider this equation for small $x_2$. Assuming no pathological behaviour of
the function $D_h^{j_vj_v}(x_1,x_2;t_0)$ near the kinematical bound
$x_1+x_2=1$, the integral on the left hand side of \eqref{valvalsum} is
dominated by contributions from the small $x_1$ region where
$D_h^{j_vj_v}(x_1,x_2;t_0)$ is largest. A factorised form for
$D_h^{j_vj_v}(x_1,x_2;t_0)$ at small $x_1,x_2$ would then result in the left hand side behaving
like $x_2^{-a_v}$ (where $x^{-a_v}$ is the small $x$ behaviour of a typical
valence sPDF).

On the other hand, the right hand side of \eqref{valvalsum} is dominated by the
$-D_h^{j+\bar{j}}(x_2;t_0)$ term. This is due to the fact that this term
receives contributions from the sea, and sea sPDFs diverge faster than valence
sPDFs at low $x$. We expect $-D_h^{j+\bar{j}}(x_2)$ to behave like $-x_2^{a_s}$
(where a typical sea sPDF behaves like $x^{a_s}$ at low $x$). The right hand
side then behaves very differently\footnote{Regge theory arguments, for example, would suggest $a_v
\simeq \frac{1}{2}$ and $a_s \simeq 1$, and `modern' global fit sPDFs show a similar trend.}
  from the left hand side, and it is
impossible to satisfy the sum rule \eqref{valvalsum} using a dPDF that
factorises at low $x_1,x_2$.

We conclude that we must abandon the possibility of factorisation into a product of sPDFs at low $x_1,x_2$ for the $D_h^{j_vj_v}(x_1,x_2;t_0)$. The fundamental origin of the second term on the right hand side of \eqref{valvalsum} which precludes the possibility of a factorised form for $D_h^{j_vj_v}(x_1,x_2;t_0)$ is of course in number effects. By `number effects' we mean the fact that finding a parton of a given type alters the probability of finding a further parton of the same type, due to the fact that the number of that parton has decreased.

The CDF and D0 results are not in contradiction with the above
conclusion, since in these experiments the vast majority of double parton scatterings
observed would have been initiated by gluons and sea quarks. The dPDFs relevant
to these partons are able to have factorised forms at low $x_1,x_2$.

At first glance, it might appear that the statement of the inadequacy of factorised forms as applied to the valence-valence distributions has already been made, in \cite{Halzen:1986ue}. However, our statement and the one in \cite{Halzen:1986ue} are really very different things. In \cite{Halzen:1986ue}, the authors argue that one should not use a factorised form for the valence-valence dPDFs at {\em large} $x_1,x_2$. The reasoning behind this is that the inaccuracies of the factorised ansatz at large $x_1,x_2$ due to the fact that it neglects momentum conservation effects are most strongly noticed in the valence-valence dPDFs, which are dominant at large $x_1,x_2$. Whilst we agree with their conclusions, we further propose that the factorised forms should not be used to describe equal flavour valence-valence dPDFs at {\em small} $x_1,x_2$, a point that is missed in \cite{Halzen:1986ue} and elsewhere.

Bearing in mind the points made above, we proceed to discuss how some input
distributions approximately obeying the sum rules might be obtained. One might initially
wonder whether it is possible to develop a framework for constructing dPDFs out
of combinations of sPDFs that does not make reference to any specific choices
for the input sPDFs (e.g. MSTW, CTEQ). Instead, it would make intelligent use
of the sum rules the sPDFs have to satisfy to ensure the dPDF sum rules were
satisfied. However, investigation into this route has revealed that a framework of
this kind does not seem to exist, even to construct dPDFs that only satisfy one of 
the two types of sum rules.

Our discussion must therefore be based around some specific set of input sPDFs. For the purposes of producing the most accurate set of dPDFs we can, it would seem sensible to use the inputs from the most recent LO fit by one of the PDF fitting collaborations. We have chosen to use a set which almost exactly corresponds to the MSTW2008 LO inputs (Equations 6-12 and the first column of Table 4 in \cite{Martin:2009iq}, with $Q_0=1$~GeV and $\alpha_s(Q_0)=0.68183$). The only differences between our inputs and those of \cite{Martin:2009iq} are that we have set the initial $s_v$ distribution to zero, and have added the following terms to the $\overline{d}$ distribution:
\begin{equation}
-148.103388x^3(1-x)^{10.8801}+500x^4(1-x)^{10.8801}
\end{equation}

These modifications have been made in order to fix the problem that the MSTW2008 LO $\overline{s}$ and $\overline{d}$ input distributions go slightly negative in some region of $x$. Even though strictly speaking these LO sPDFs should never go negative, the deviations below zero observed in the MSTW2008 LO $\overline{s}$ and $\overline{d}$ inputs are perhaps tolerable in single scattering calculations due to their small size ($\overline{s},\overline{d} > -0.0005$). However, we must insist on using sPDFs which are strictly non-negative when expressed in the `human' flavour basis\footnote{The `human' flavour basis is the one in which the parton index $i = g,u,\bar{u}$, etc.} to build our input dPDFs. We can explain why this has to be the case by considering the dPDFs in the `double human' basis in which at least one flavour index corresponds to an sPDF which goes negative. Like all LO dPDFs in the `double human' basis, they cannot go negative (due to their interpretation as a probability). If we use a pseudo-factorised prescription to construct the dPDFs, then these dPDFs will go very seriously negative where the sPDF in one direction takes small negative values, and the sPDF in the other becomes large and positive. We therefore require strictly non-negative input sPDFs.

We can identify two key features that we would like to build in to our set of input dPDFs. These are the following:

\begin{enumerate}
    \item The dPDFs should be suppressed below factorised values near the kinematical bound (i.e. the line $x_1+x_2=1$) due to phase space considerations. 

    \item Terms should be added/subtracted from certain dPDFs to take account of number effects. 
\end{enumerate}

Let us begin by discussing how the first requirement might be incorporated. In the early papers \cite{Humpert:1983pw,Humpert:1984ay,Halzen:1986ue,Goebel:1979mi}, a common $(1-x_1-x_2)$ suppression factor multiplying all of the dPDFs was advocated. This was motivated by arguments based on the recombination model of \cite{Das:1977cp}, or the Kuti-Weisskopf model of \cite{Kuti:1971ph}. More recently \cite{Korotkikh:2004bz}, it has been suggested that a higher power of $(1-x_1-x_2)$, such as $(1-x_1-x_2)^2$, might be appropriate. With the benefit of knowledge of the sum rules, we can see that neither of these alternatives is entirely satisfactory. To illustrate this, let us just consider the momentum sum rule for the moment (which is the relevant rule with regards to phase space considerations), and let us consider the lines $x_1=0$ and $x_2=0$. Along these lines, all momentum sum rules are perfectly satisfied using factorised dPDFs, whilst dPDFs including a $(1-x_1-x_2)$ or $(1-x_1-x_2)^2$ factor violate the sum rules badly.

Thus a $(1-x_1-x_2)^n$ factor alone multiplying all of the dPDFs suppresses the functions rather too severely near the lines $x_1=0$ and $x_2=0$, and it would seem that a phase space factor which approached $1$ near these lines would be more desirable. We can actually make sense of this from an intuitive point of view. The phase space suppression factor is inserted to take account of the fact that finding a parton with $x=x_1$ reduces the probability of finding another parton with $x=x_2$ if $x_1+x_2$ is close to $1$. One would expect a much smaller reduction if $x_1$ were small and $x_2$ were large than if both $x_1$ and $x_2$ were large, even if the sum of $x_1$ and $x_2$ was the same in both cases. Indeed, one would anticipate that the reduction should tend to zero as $x_1$ (or $x_2$) tended to zero -- that is, the phase space factor should approach $1$ as one approaches the lines $x_1=0$ and $x_2=0$.

Here, we continue to follow the tradition set by previous papers in that we have attempted to apply a universal phase space factor to all of the dPDFs. Use of a (positive) universal phase space factor has the advantage that it is guaranteed to produce positive double human basis dPDFs. However, instead of using $(1-x_1-x_2)^n$ alone, we tried the following as a `first guess' for the phase space factor $\rho$, motivated by the above discussion:
\begin{equation} \label{pfact1}
\rho(x_1,x_2) = (1-x_1-x_2)^n(1-x_1)^{-n}(1-x_2)^{-n}
\end{equation}

\begin{figure}
\centering
\includegraphics[scale=0.9]{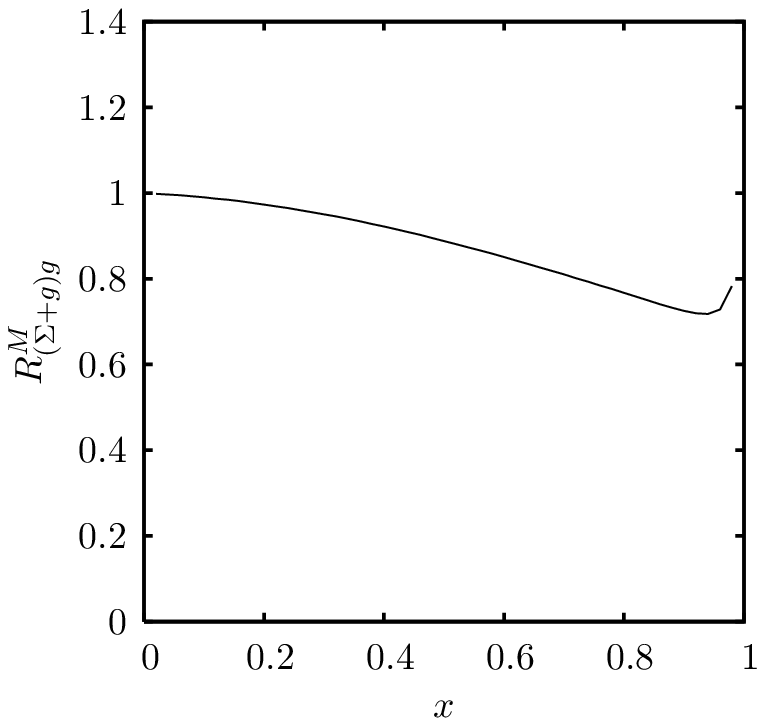}
\includegraphics[scale=0.9, trim = 1cm 0 0 0]{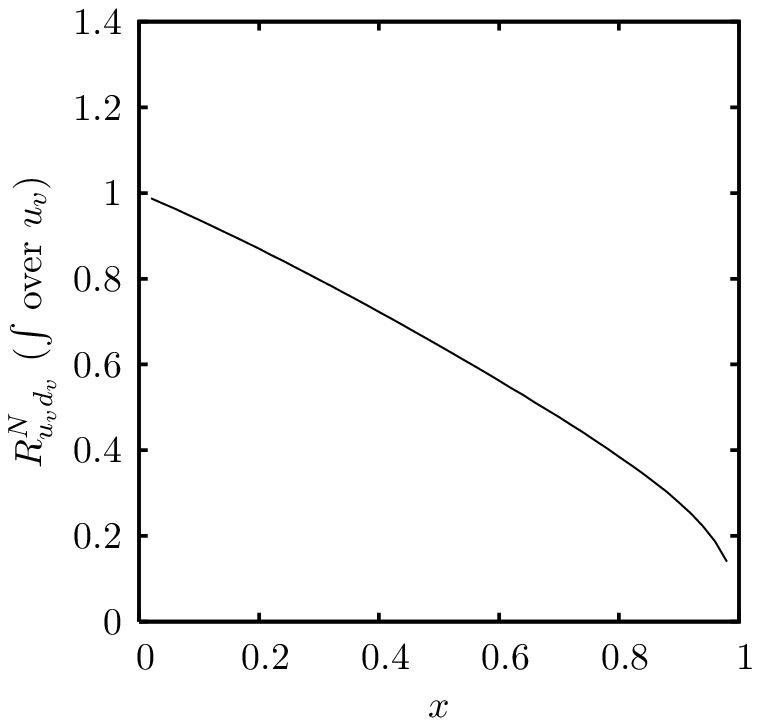}
\caption{\label{fig:Simplephasefact} Sum rule ratios for the $(\Sigma+g)g$ momentum and $u_vd_v$ (integrating over $u_v$) number sum rules, when the phase space factor is as given in \eqref{pfact1} with $n=2$.}
\end{figure}

Following the more recent work by Korotkikh and Snigirev \cite{Korotkikh:2004bz}, we choose $n$ to be $2$. This choice of phase space factor gave dPDFs which satisfied the momentum sum rules reasonably well. In the left panel of Fig.~\ref{fig:Simplephasefact}, we plot the `sum rule ratio' with this phase space factor for the particular example of the $(\Sigma+g)g$ momentum sum rule -- the sum rule ratios for the other momentum sum rules exhibit very similar behaviour. The sum rule ratio for a particular sum rule and set of dPDFs is defined as the sum rule integral calculated using the dPDFs divided by the sPDF quantity it should be equal to. It is a function of an $x$ variable, and measures how well the dPDFs satisfy the given sum rule -- the closer the ratio is to $1$ over the full $x$ range, the better the dPDFs satisfy the sum rule.

On the other hand, the dPDF number sum rules are not particularly well satisfied by this prescription (this is illustrated in the right panel of Fig.~\ref{fig:Simplephasefact}). This is true even for those dPDFs which are involved in a number sum rule but which are not affected by number effects -- e.g. $u_vd_v$. For these dPDFs, the phase space factor alone should be sufficient to cause the dPDFs to satisfy the relevant number sum rules -- thus our first guess is not fully satisfactory. We have discovered that a slight adjustment to the form \eqref{pfact1} resolves this problem. Let us allow the phase space factor to depend on the parton indices $i,j$ on the dPDF such that (prior to adjustments relating to point 2 above) the input dPDFs are constructed according to:
\begin{equation}
D^{ij}_h(x_1,x_2;t_0) = D^{i}_h(x_1;t_0)D^{j}_h(x_2;t_0)\rho^{ij}(x_1,x_2)
\end{equation}

We now define $\rho^{ij}(x_1,x_2)$ as follows:
\begin{equation}\label{pfact2}
\rho^{ij}(x_1,x_2)=(1-x_1-x_2)^2(1-x_1)^{-2-\alpha(j)}(1-x_2)^{-2-\alpha(i)}
\end{equation}

\noindent where:
\begin{eqnarray}
\alpha(i) = \begin{cases} 0 & \text{if $i$ is a sea parton} \\
0.5 & \text{if $i$ is a valence parton}
\end{cases}
\end{eqnarray}

If either $i$ and/or $j$ contain both valence and sea contributions, then one should construct the dPDF by taking the factorised product, splitting it into sets of terms corresponding to valence-valence, valence-sea, sea-sea, etc., and then applying the appropriate phase space factor to each set of terms. Note that the phase space factor is no longer universal, but is nearly so -- it turns out that this prescription is guaranteed to produce positive human basis dPDFs provided all the valence sPDFs are positive, which is the case for the set we have chosen.

\begin{figure}
\centering
\includegraphics[scale=0.9]{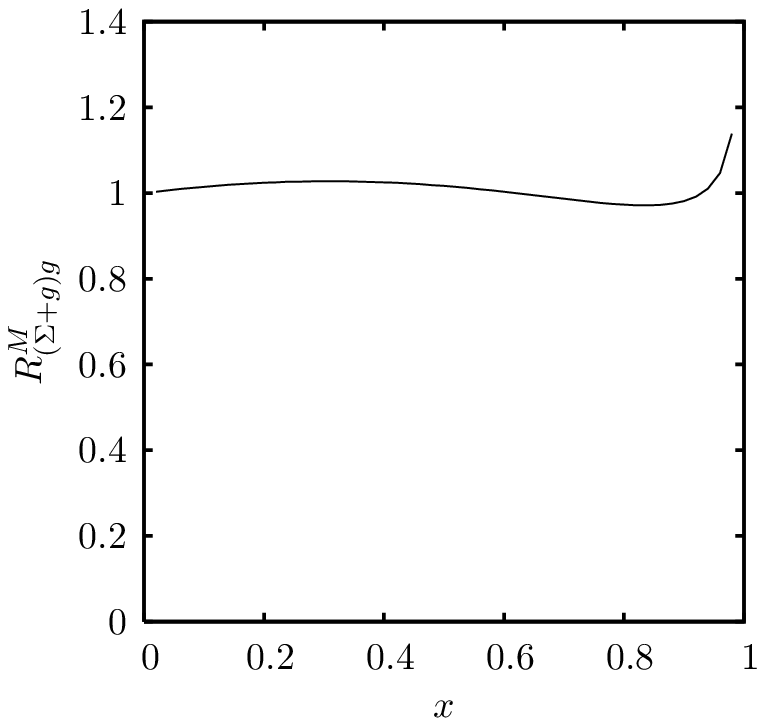}
\includegraphics[scale=0.9, trim = 1cm 0 0 0]{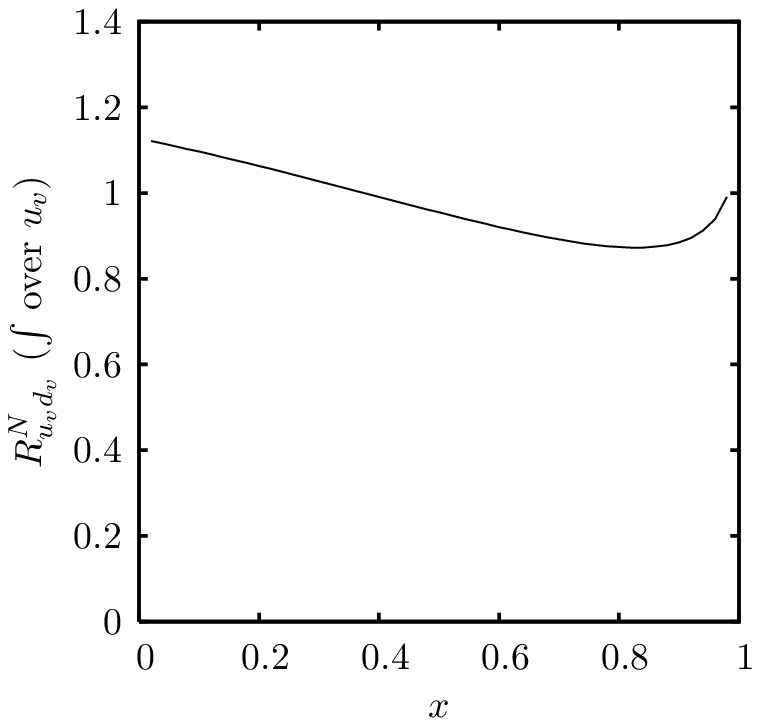}
\caption{\label{fig:Improvedphasefact} The same sum rule ratios as in Fig.~\ref{fig:Simplephasefact}, but this time plotted with the phase space factor as in \eqref{pfact2}.}
\end{figure}

With the choice \eqref{pfact2}, the dPDFs involved in number sum rules but which are not affected by number effects satisfy their sum rules to a much better degree. It also turns out that once we have included terms to take account of number effects (described shortly), insertion of phase space factors according to \eqref{pfact2} into dPDFs affected by these effects similarly improves the degree to which these dPDFs satisfy the sum rules. What is more, the momentum sum rules are much better satisfied when one uses \eqref{pfact2} rather than \eqref{pfact1}. Illustration of some of these points for some representative dPDF cases, as well as an exposition of the extent to which we satisfy the sum rules with this choice of phase space factor, is given in Fig.~\ref{fig:Improvedphasefact}.

Having found a satisfactory phase space factor, we proceed to discuss how the second required feature in the list above -- namely the incorporation of number effects -- might be achieved in our input dPDFs. We have seen that number effects are particularly important for equal flavour valence-valence dPDFs, and we shall outline how suitable inputs for this particular type of dPDF may be constructed shortly. However, number effects can also in principle have an impact on any other dPDF for which the same parton type appears in both parton indices. Since there are only a finite number of valence up and down quarks in the proton (as opposed to an infinite number of sea quarks and gluons), one might anticipate number effects relating to these valence quarks to be most important. We now discuss how these effects can be included in dPDFs which `contain' an up and/or a down valence combination in both of their parton indices (e.g. $u_+u_v$, $d_+d_+$, where $i_+ \equiv i+\overline{i}$). 

An example of such a distribution would be the $u_+u_+$ distribution, since $u_+u_+ = (u_v+2u_s)(u_v+2u_s)$, where $u_s=\overline{u}$. Consider the ways in which one can pick two up flavour partons (either quarks or antiquarks) from the proton. Either one can pick two sea partons, or one can pick a sea parton and a valence quark (in either order), or one can pick two valence quarks -- these possibilities of course correspond to the different terms in the expansion of $(u_v+2u_s)(u_v+2u_s)$. Factorised terms multiplied by phase space factors are reasonable for all possibilities apart from the two valence option, where it would seem important to take account of the fact that removing a valence up halves the probability to find another. At a crude level we can incorporate this fact by using a term which is equal to half of the naive `factorised $\times$ phase space factor' guess for the valence-valence term. We can think of this adjustment in another way, and say that we incorporate number effects in the $u_+u_+$ distribution by subtracting the following term from our initial `factorised $\times$ phase space factor' construct:
\begin{equation} \label{upCVNcorrection}
\dfrac{1}{2}D^{u_v}_h(x_1;t_0)D^{u_v}_h(x_2;t_0)\rho^{u_vu_v}(x_1,x_2)
\end{equation}

\begin{figure}
\centering
\includegraphics[scale=0.9]{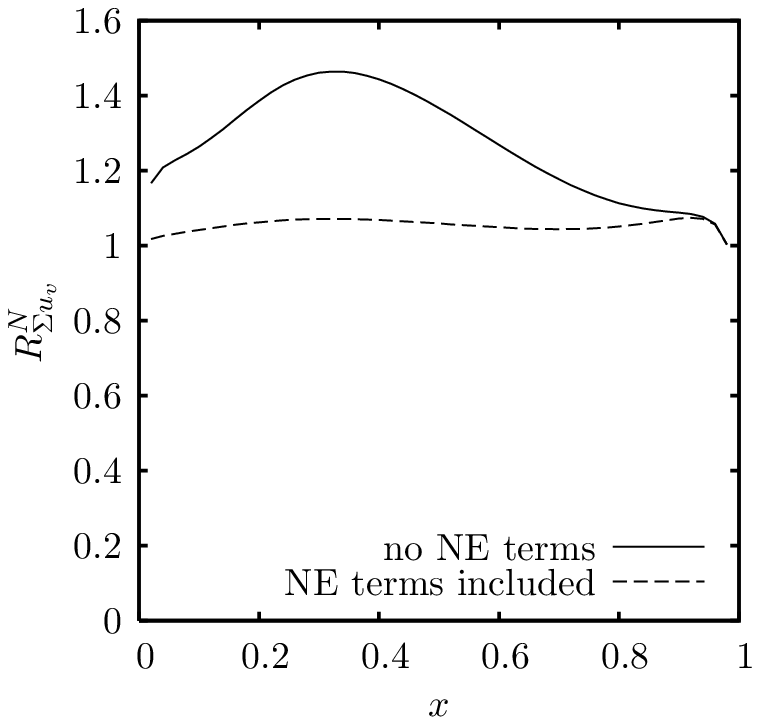}
\includegraphics[scale=0.9]{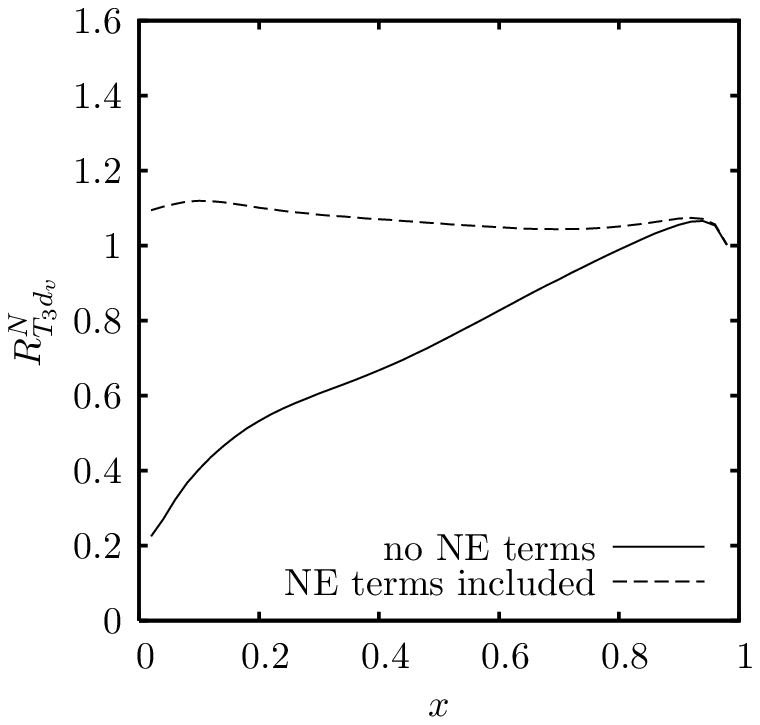}
\caption{\label{fig:Numberterms} The effect of adding number effect (NE) terms on the sum rule ratios for the $u_v \Sigma$ and $d_v T_3$ number sum rules.}
\end{figure}

Generalising this argument, we observe that a dPDF which contains $n$ times the up valence--up valence combination in its parton indices must have $n$ times the term \eqref{upCVNcorrection} subtracted from it to take account of number effects. Similarly, a distribution which contains $n$ times the down valence-down valence combination in its parton indices must have $n$ times $D^{d_v}_h(x_1;t_0)D^{d_v}_h(x_2;t_0)\rho^{d_vd_v}(x_1,x_2)$ subtracted from it. Note in this case that we must remove the naive $d_vd_v$ term entirely because there is no chance of finding two valence down quarks in the proton. Fig.~\ref{fig:Numberterms} shows how inclusion of the number effect terms improves the extent to which dPDFs satisfy number sum rules, for a few sample cases. 

We now turn our attention to the construction of some equal flavour valence-valence dPDFs approximately satisfying the sum rules. The flavours we must be concerned about here are up, down, {\em and strange}. Note that the $s_vs_v$ distribution is not zero with the given set of input sPDFs, even though the $s_v$ sPDF is zero. The sum rule for this dPDF reads:
\begin{equation} \label{svsvrule}
\int_0^{1-x_2}dx_1D_h^{s_vs_v}(x_1,x_2;t_0) =
-D_h^{s_+}(x_2;t_0)
\end{equation}

Since the MSTW 2008LO $s_+$ input is nonzero, the right hand side of \eqref{svsvrule} is nonzero, and consequently the $s_vs_v$ dPDF cannot be zero. We can explain why the $s_vs_v$ distribution should be nonzero by expanding the combination into double human basis pairs -- $s_vs_v = ss-s\bar{s}-\bar{s}s+\bar{s}\bar{s}$. We expect the probability to find an $s\bar{s}$ pair to be higher than that to find an $ss$ or $\bar{s}\bar{s}$ pair due to number effects. Given that one has found a strange (antistrange) in the proton, the probability to find a further strange (antistrange) is reduced, whilst that to find an antistrange (strange) in addition remains the same.

In order to construct satisfactory distributions for these three flavour types, we imagine that there exists a scale $\tilde{t}<t_0$ at which only the three valence quarks in the proton may be resolved, and all sea distributions are zero. The sea distributions at $t_0$ are then generated dynamically by DGLAP evolution between $\tilde{t}$ and $t_0$. This idea has previously been put forward in \cite{Parisi:1976fz,Vainshtein:1976kd,
Novikov:1976dd,Gluck:1977ah,Gluck:1988xx}, in which it was investigated whether the possibility exists to fit deep inelastic scattering data using only $u_v$ and $d_v$ inputs at a fitted low scale $\tilde{t}$. As it turns out, one cannot achieve a fully satisfactory fit of data using this approach, as is admitted in \cite{Gluck:1989ze}. However, since we shall only use this idea very loosely in what follows, this point is not of great concern to us.

At the scale $\tilde{t}$, the only equal flavour valence-valence dPDF which can be nonzero is the $u_vu_v$ distribution, as there is no possibility of finding two down or strange partons (be they quarks or antiquarks) at this scale. A suitable ansatz for the $u_vu_v$ at $\tilde{t}$ is a product of $u_v$ sPDFs multiplied by a phase space factor $\tilde{\rho}$ appropriate at the scale, and divided by two to take account of valence-valence number effects:
\begin{equation} \label{valonuv}
D_h^{u_vu_v}(x_1,x_2;\tilde{t}) = 
\dfrac{1}{2}D_h^{u_v}(x_1;\tilde{t})D_h^{u_v}(x_2;\tilde{t})
\tilde{\rho}^{u_vu_v}(x_1,x_2)
\end{equation}

One can straightforwardly verify that the above forms for the equal flavour valence-valence dPDFs are consistent with the number sum rules at this scale. Now let us consider how the dPDFs change as we evolve from $\tilde{t}$ to $t_0$ under \eqref{dbDGLAP}. The first two sets of terms on the RHS of \eqref{dbDGLAP} will mainly serve to take \eqref{valonuv} into its equivalent at $t_0$ (and leave the other equal flavour valence-valence distributions zero). However, the final set of `sPDF feed' terms results in an extra contribution appearing in each equal flavour valence-valence dPDF. Only the $-j\bar{j}-\bar{j}j$ component of an equal flavour valence-valence combination receives nonzero sPDF feed contributions during evolution ($g \to j\bar{j}$ contributions). Therefore, the sPDF feed for an equal flavour valence-valence dPDF is the following:
\begin{equation}
-2\dfrac{\alpha_s(t)}{2\pi}D^{g}_h(x_1+x_2;t)\dfrac{1}{x_1+x_2}P_{qg}\left(\dfrac{x_1}{x_1+x_2}\right)
\end{equation}

The splitting function $P_{qg}$ is not a very strong function of its argument (only varying between $\frac{1}{2}$ and $\frac{1}{4}$). This means that, roughly speaking, we can take the sPDF feed term for the equal flavour valence-valence distributions as being a function of $(x_1+x_2)$. If we then ignore the subsequent effect of the first two sets of terms on the RHS of \eqref{dbDGLAP} on the sPDF feed contributions, then we expect the sum total sPDF feed contribution to each valence-valence dPDF at $t_0$ to be a function of $(x_1+x_2)$ only:
\begin{align} \label{EFVVdist}
D_h^{j_vj_v}(x_1,x_2;t_0) = \dfrac{N_{j_v}-1}{N_{j_v}}D_h^{j_v}(x_1;t_0)D_h^{j_v}(x_2;t_0)
\rho^{j_vj_v}(x_1,x_2)-2g^{j\bar{j}}(x_1+x_2;t_0)
\end{align}

We shall refer to the function $g^{j\bar{j}}(x_1+x_2;t_0)$ as the $j\bar{j}$ correlation term, as it represents the `nonfactorised' part of the $j\bar{j}$ (or $\bar{j}j$) distribution which is built up from correlation-inducing sPDF feed contributions. How should we decide on the form of this function for a particular choice for the flavour $j$? We can answer this question by using the number sum rule that \eqref{EFVVdist} must satisfy, which we shall write here as:
\begin{equation} \label{EFVVnumsum}
\int_0^{1-x_2}dx_1D_h^{j_vj_v}(x_1,x_2;t_0) = 
(N_{j_v}-1)D_h^{j_v}(x_2;t_0)-2D_h^{\bar{j}}(x_2;t_0)
\end{equation}

The first term on the RHS of \eqref{EFVVdist} integrates to give approximately the first term on the RHS of \eqref{EFVVnumsum}. The $-2g^{j\bar{j}}(x_1+x_2;t_0)$ must therefore integrate to give the second term on the RHS of this equation:
\begin{equation} \label{ginteq}
-2\int_0^{1-x_2}dx_1g^{j\bar{j}}(x_1+x_2;t_0) = -2D_h^{\bar{j}}(x_2;t_0)
\end{equation}

This is an integral equation with a unique solution, and it is straightforward to show that the solution is the following:
\begin{equation} \label{gsoln}
g^{j\bar{j}}(x;t_0) = -\dfrac{\partial D_h^{\bar{j}}(x;t_0)}{\partial x}
\end{equation}

\begin{figure}
\centering
\includegraphics[scale=0.9]{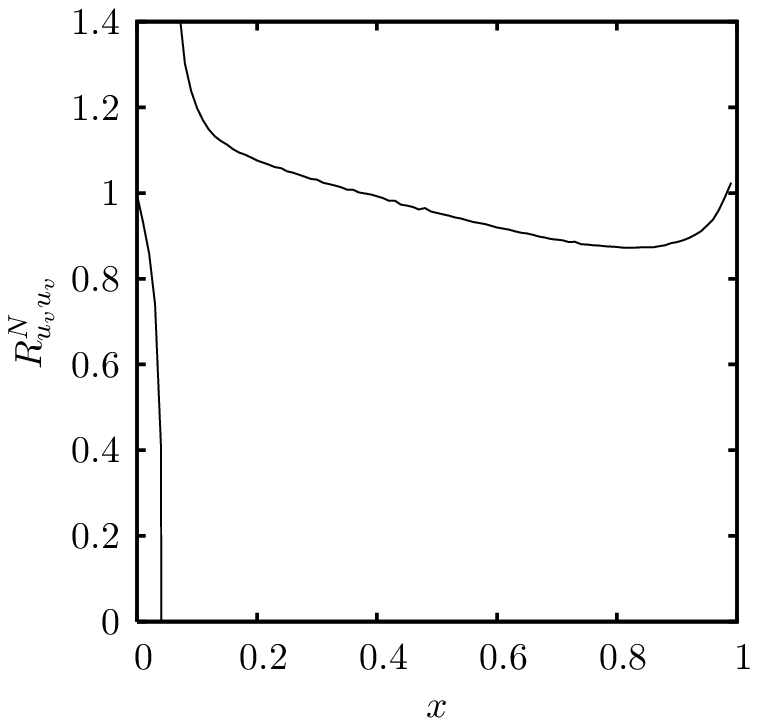}
\includegraphics[scale=0.9]{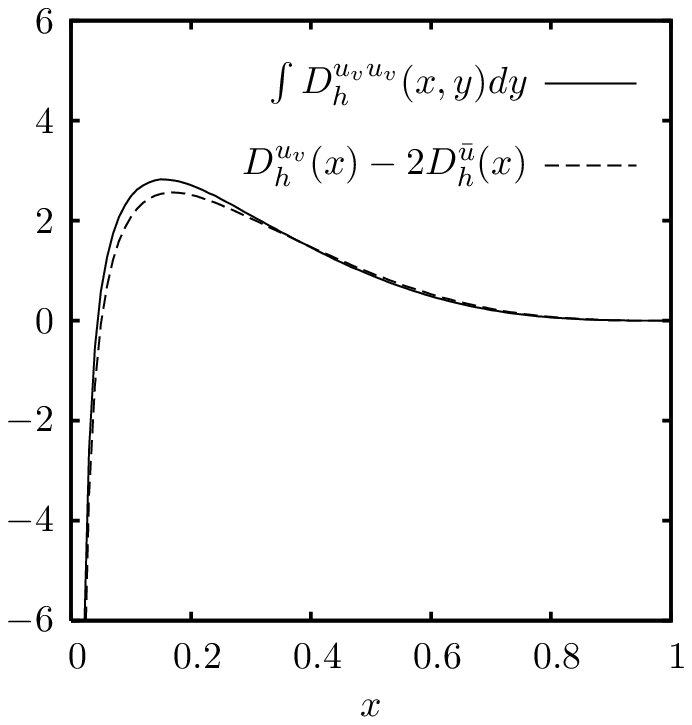}
\caption{\label{fig:uvuv} {\em Left panel}: The sum rule ratio for the $u_vu_v$ number sum rule when $D_h^{u_vu_v}$ is constructed according to \eqref{EFVVdist} and \eqref{gsoln}. The ratio is close to $1$ over most of the range of $x$, except near $x=0.05$ where it diverges violently. This appears to indicate that the sum rule is being badly violated near $x=0.05$. \\
{\em Right panel}: The $u_vu_v$ sum rule integral plotted against the sPDF quantity it should be equal to. This plot reveals that the divergence in the sum rule ratio is caused by the integral curve slightly missing a zero in the sPDF quantity, and is not serious in practice.}
\end{figure}

Our proposed form for the input equal flavour valence-valence distributions is therefore \eqref{EFVVdist} with $g^{j_vj_v}$ given by \eqref{gsoln}. Clearly the $d_vd_v$ and $s_vs_v$ number sum rules will be perfectly satisfied using this form. Fig.~\ref{fig:uvuv} shows how well the $u_vu_v$ sum rule is satisfied.

Unfortunately, with this choice for the equal flavour valence-valence dPDFs, the $\bar{u}\bar{u}$, $\bar{d}\bar{d}$, $ss$ and $\bar{s}\bar{s}$ dPDFs all go negative. Naively, one might view this as arising because the forms we have used for the equal flavour valence-valence dPDFs are in some way unsatisfactory. However, instead we observe that it occurs because we have omitted an important term in our above treatment of the $j_+j_+$ distributions. Since these distributions contain the parton combination $j\bar{j}+\bar{j}j$ that also appears in the $j_vj_v$ distribution with the opposite sign, the $j_+j_+$ receive the same sPDF feed contributions as the $j_vj_v$ during evolution, but with the opposite sign. Thus for consistency each $j_+j_+$ distribution should have an extra term added onto it equal to {\em plus} $2g^{j\bar{j}}(x_1+x_2;t_0)$. With this alteration, all double human basis dPDFs are again positive, and we see little adverse effect on the extent to which the sum rules involving $j_+j_+$ distributions are satisfied.

\begin{figure}
\centering
\includegraphics[scale=0.9]{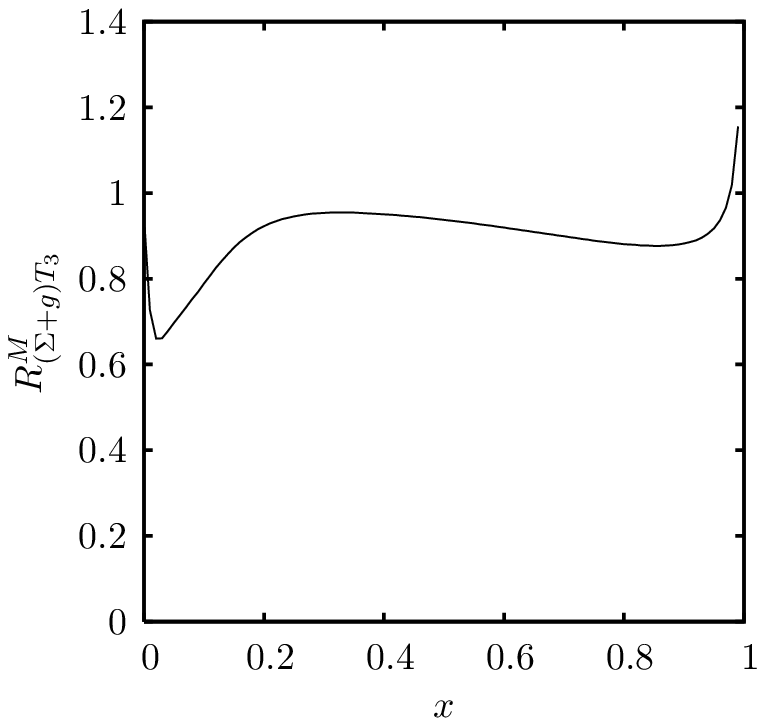}
\hspace{0.5cm}
\includegraphics[scale=0.9]{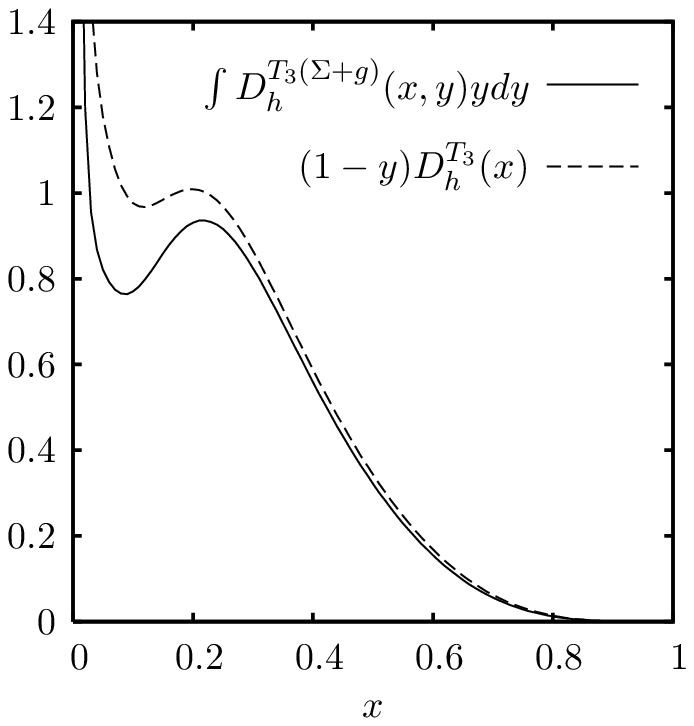}
\caption{\label{fig:SgT3} {\em Left panel}: The sum rule ratio for the $(\Sigma+g)T_3$ momentum sum rule, plotted using the fully constructed set of input dPDFs. \\
{\em Right panel}: The $(\Sigma+g)T_3$ momentum sum rule integral plotted against the sPDF quantity it should be equal to.}
\end{figure}

Having now completed our description of how we constructed some suitable input dPDFs, we conclude our discussion with a short summary of how well the dPDFs satisfy the complete set of sum rules. In the context of the double human basis, the sum rule ratios are all within $25\%$ of 1 for $x\lesssim 0.8$. Above this value, the sum rules are not obeyed so well -- however the values of the PDFs are tiny at these $x$ values, so large/small sum rule ratio values at these $x$ values are not in practice too great a problem. In the double evolution basis the story is the same, barring trivial divergences due to the sum rule integral slightly missing a zero in the sPDF quantity it should be equal to. The one exception to this is the case of the $T_3(\Sigma+g)$ momentum sum rule. The sum rule ratio for this sum rule, plotted in the left panel of Fig.~\ref{fig:SgT3}, plunges to 0.65 around $x=0.02$. This possibly looks worse than it is -- if one plots both the integral and the sPDF quantity it should be equal to (right panel of Fig.~\ref{fig:SgT3}), then one notices that the dip in the sum rule ratio is due to the integral slightly overestimating a dip in the sPDF quantity in a region where the sPDF quantity is rather small. Furthermore, it is unlikely that the particular combination $T_3(\Sigma+g)$ will be directly accessed by any scattering processes at the LHC. Consequently we are prepared to accept the large deviation from $1$ in the $T_3(\Sigma+g)$ sum rule ratio.

\section{Numerical Solution of the Double DGLAP Equation} \label{sec:method}

There exist several options for the broad numerical method to use to integrate
the dDGLAP equations. One could choose to adapt either the direct $x$ space or
Mellin transform methods which are commonly used to numerically integrate the
sDGLAP equation (see, for example, \cite{Botje:2009bj,Salam:2008qg} for
routines using the $x$ space method for solution of the sDGLAP equation, and
\cite{Vogt:2004ns} for  a routine using the Mellin transform method).
Alternatively, one could develop a numerical method based on the explicit
solution of the dDGLAP equation in terms of sPDFs \eqref{dbDGLAPsoln}. This is
the approach that has been preferred in the previous numerical treatments of
the subject \cite{Korotkikh:2004bz, Cattaruzza:2005nu}. Here we adopt an $x$
space method. This has the advantages that it is conceptually simple, is
flexible enough to take the inputs described in Section~\ref{sec:sumrulesinput}
with no problems, and is competitive in efficiency with the other methods in
the context of the dDGLAP equation. It also has the advantage over the
`explicit solution' method in that the $D_i^j(x;t)$ Green's functions, which
are difficult to calculate numerically to a sufficient degree
of accuracy, do not feature.

\subsection{The dDGLAP Evolution Program}
\label{subsec:program}

Our program solves the dDGLAP equation \eqref{dbDGLAP} directly using a grid in
$x_1,x_2$ and $t$. We choose the spacing of the grid points in $t$ to be linear
-- this is the `natural' choice, and it is adopted in a number of sDGLAP
$x$-space routines (e.g. \cite{Botje:2009bj, Salam:2008qg}). In the $x_1$ and
$x_2$ directions, the points are taken to be evenly spaced in the variable
$u=\ln(\tfrac{x}{1-x})$, with equal numbers of points in the $x_1$ and $x_2$
directions ($600$ for the grids of \cite{HepForgePage}). 
This gives a spacing uniform in $\ln(x)$ in the small $x$ regions and directions in which
the dPDF is diverging rapidly, and a linear spacing in larger $x$ regions
and directions in which the variation of the dPDF is slower.
The boundary of the grid in ($x_1$,$x_2$) space is defined by the
lines $x_1=x_{\rm min}$, $x_2=x_{\rm min}$, $x_1=1-x_{\rm min}$, $x_2=1-x_{\rm min}$, and
$x_1+x_2=1$ (the kinematical boundary), with a default $x_{\rm min} = 10^{-6}$. The methods we 
use for the numerical integration
of the first two terms on the right hand side of the dDGLAP equations are described in the Appendix.

The final set of terms in the dDGLAP equation (the `sPDF feed' terms) are
obtained at a given $t$ by numerically evolving the sDGLAP equations
contemporaneously with the dDGLAP equations. The grid used for the sDGLAP
evolution is the similar to that used for the dDGLAP evolution. The only
difference is that it extends in just one $x$ direction, between $x_{\rm min}$ and
$(1-x_{\rm min})$. For consistency, the sPDF inputs used are the MSTW2008LO inputs.

Given the structure of the dDGLAP equation, the dDGLAP evolution routine
requires the values of the sPDFs at $x$ values of the form $x_i+x_j$, where
$x_i$ and $x_j$ are two $x$ values on the uniform in $\ln(x/(1-x))$ grid. With
the grid used, it is clear that $x_i+x_j$ does not also lie on the grid, so
interpolation has to be used to obtain the sPDF values required. Away from the
edges of the sPDF $x$-grid, natural cubic spline interpolation based on the
sPDF values at the nearest four grid points is used, whilst linear
interpolation is used at the edges.

The program uses the `double evolution' basis introduced in Section
\ref{sec:sumrules} as its internal basis for the evolution of the dPDFs. Use of
this basis for the evolution is advantageous because the dDGLAP equations
become in some sense `minimally coupled' in this basis. Out of the $91$
equations, $66$ are rendered diagonal at LO using this basis (i.e. rate of
change of $D_h^{ij}$ with $t$ is given only by the two integral terms involving
$D_h^{ij}$, with no nonzero sPDF feed terms). The remaining equations have very
few terms on the RHS (two terms in each integral term plus one sPDF feed term).
The use of this basis makes the coding in of the dDGLAP equations manageable.
%The program is equipped with a subroutine which can convert the dPDFs back into
%the normal flavour basis (i.e. the one in which the flavour indices $i,j =
%g,u,\bar{u}$, etc.) at the end of the evolution if this is required. 

Stepwise evolution in $t$ is carried out by a fourth-order Runge-Kutta method.
The evolution begins at a scale $t_0$ equal to that at which the input
distributions are defined ($Q_0^2 = 1$~GeV$^2$ with the MSTW2008LO inputs). 
The final scale obtained in the evolution $t_{f}$ and the
number of Runge-Kutta steps used to reach this scale $N_t$ may be specified by
the user. To produce the grids of \cite{HepForgePage}, $120$ points were used in the $t$
direction.

\subsection{Flavour Number Schemes}
\label{subsec:flavour}

Our program has the potential to perform the evolution using either a fixed or
(zero mass) variable flavour number scheme, with $n_f$ fixed at $3, 4,5$ or $6$ in the FFNS,
or potentially varying from $3\to 6$ in the ZM-VFNS. The scheme can be determined
by the user via the variables \v{LGMCSQ}, \v{LGMBSQ} and \v{LGMTSQ} which are equal to the
thresholds in $t$ at which the charm, bottom and top flavours become active
respectively. For a FFNS of given $n_f$, \v{LGMCSQ}, \v{LGMBSQ} and \v{LGMTSQ} should be
set appropriately either above $t_0$ or below $t_f$ (e.g. for a FFNS with
$n_f=5$, set $\v{LGMCSQ}<t_0$ ,$\v{LGMBSQ}<t_0$ and $\v{LGMTSQ}>t_f$). For a ZM-VFNS, at 
least one of \v{LGMCSQ}, \v{LGMBSQ} and \v{LGMTSQ} must lie in between $t_0$ and $t_f$.
It should be noted that to produce the grids of \cite{HepForgePage}, the program was run
under a ZM-VFNS with $n_f$ varying between 3 and 5. The variables \v{LGMCSQ} and \v{LGMBSQ} were
set according to the values of $m_c$ and $m_b$ preferred by MSTW 
-- $1.40\rm{~GeV}$ and $4.75\rm{~GeV}$
respectively.

Prior to the evolution, the program compares \v{LGMCSQ}, \v{LGMBSQ} and \v{LGMTSQ} with
$t_0$ and $t_f$. Depending on the results of this, it splits the full evolution
from $t_0$ and $t_f$ into up to four intervals, each with a different value of
$n_f$. The total number of integration steps in $t$, $N_t$, is divided up
amongst these intervals roughly in proportion to the interval sizes in $t$.

In each interval, the strong coupling constant $t$ is calculated according to
the LO analytic form:
\begin{align}
\alpha_S(t) &= \dfrac{\alpha_S(t')}{1+\alpha_S(t')b(t-t')}; \hspace{10pt} b
\equiv \dfrac{33-2n_f}{12\pi} .
\end{align}
The quantity $t'$ corresponds to the value of $t$ at the beginning of the
interval. In the first interval, the boundary value of the strong coupling
constant, $\alpha_S(t')$, is taken to be the initial value specified by the
user $\alpha_S(t_0)$. In later intervals it is chosen to ensure continuity in
$\alpha_S$, which is the appropriate matching condition at LO
\cite{Ellis:1996ws}.

\subsection{Accuracy of the Program} \label{sec:accspeed}

We wish to get a rough estimate of the error in the dPDF values at $Q$ introduced
by numerical evolution with $N_x$ points in each $x$ direction, and $N_t$ points in
the $t$. To do this, one might propose doing an evolution with twice as many points
in each direction, and then taking the error in the original dPDFs at $Q$ as being the 
absolute difference between the dPDF values produced by the two evolutions.
Unfortunately, we cannot perform this procedure for the values of $N_x$ and $N_t$ used
to produce the grids in \cite{HepForgePage} (600 and 120). This is because doubling the 
number of $x$ points in this case causes the program to require far more RAM than a 
typical modern machine can provide. Instead, we show here that the accuracy of the 
program is reasonable even when $N_x$ and $N_t$ take on the smaller values of $150$
and $10$ respectively -- we then know that the accuracy of the procedure with $N_x=600$
and $N_t=120$ should be very good.

\begin{figure}
\centering
\includegraphics{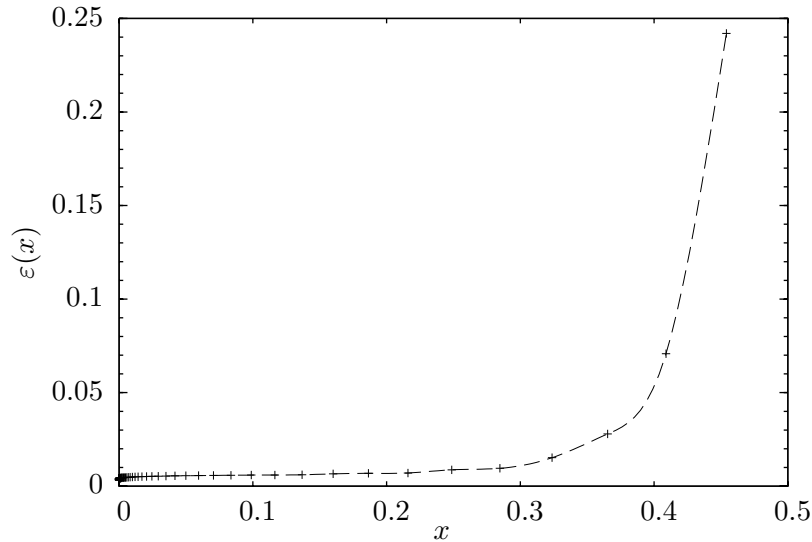}
\caption{\label{fig:accuracyplot} An estimation of the numerical error when one performs an evolution from $Q=1$~GeV to $Q=100$~GeV using a grid with $150$ points in each $x$ direction, and $10$ in the $t$. The error values plotted are those in the $gg$ dPDF along the line $x_1=x_2=x$.}
\end{figure}

We perform the error estimation evolution from $Q_0 = 1$~GeV to $Q_f = 100$~GeV. In 
Fig.~\ref{fig:accuracyplot}, the fractional error in the distribution $D_h^{gg}$ 
along the sample line $x_1 = x_2 = x$ as calculated by the above method is 
plotted. That is, we plot:
\begin{equation}
\varepsilon(x;Q_f) \equiv \dfrac{\mid D_{h}^{gg}(x,x,Q_f)_{N_x=150,N_t=10} -
D_{h}^{gg}(x,x,Q_f)_{N_x=300,N_t=20} \mid}
{D_{h}^{gg}(x,x,Q_f)_{N_x=300,N_t=20}} .
\end{equation}

We choose to look at $D_h^{gg}$ because
this is one of the dPDFs which should be calculated least accurately by an
evolution routine. As
expected, the error increases as one approaches the kinematical bound due to
the fact that less $x$ points are used in the evolution integrations for the
dPDF values closer to the bound. We see that the error is small in the crucial
small $x$ region -- less than $1\%$ for $x\lesssim 0.3$, and less than $6\%$ for
$x\lesssim 0.4$. The error becomes large as one approaches $x=0.5$, but since
this region is not likely to be important in applications at the LHC (which probes $x_1,x_2
\lesssim 0.1$), this is not a major problem. The graph indicates that even with 
$N_x=150$ and $N_t=10$ the numerical evolution to LHC scales introduces errors
which are less than $1\%$ for $x_1<0.3, x_2<0.3$, and less than $6\%$ for 
$x_1<0.4, x_2<0.4$.

\section{Properties of the dPDFs} \label{sec:properties}

We have seen that there are two ways to improve on using simple products of sPDFs as the dPDFs at the (high) scale $Q$. First, one can use dDGLAP evolution to obtain the dPDFs at $Q$, with a reasonable choice of dPDFs at a low scale $Q_0$ used as the starting point for the evolution. Second, one can use improved inputs at the low scale $Q_0$, which take account of momentum and number effects. In this section, we describe and illustrate the extent to which introducing these improvements changes the dPDFs at the scale $Q$. 

The large number of dPDFs precludes the possibility of discussing them all. Instead, we choose to focus on a small number of parton pairings which should be important in double scattering processes at the LHC, and which in some sense might be considered to form a representative set. These are the $uu$, $u\bar{u}$, $ug$ and $gg$ pairings. Note that we have a dPDF for which our input form contains a valence number effect term in this set (the $uu$), and a distribution for which our input contains a $j\bar{j}$ correlation term (the $u\bar{u}$). Furthermore, we see that the set covers all types of sPDF feed term that can appear in dDGLAP evolution.

For the purposes of making concrete comparisons between different methods of obtaining the dPDFs at a high scale $Q$, we also need to make a specific choice for $Q$. Except where otherwise stated, we make the reasonable choice $Q = 100$~GeV ($\sim M_W, M_Z$, for example). At the scale $Q$, we only look at the dPDF values along the line $x_1=x_2$ -- this allows us to produce easily readable 2D plots.

\begin{figure}
\centering
\includegraphics[scale=0.9]{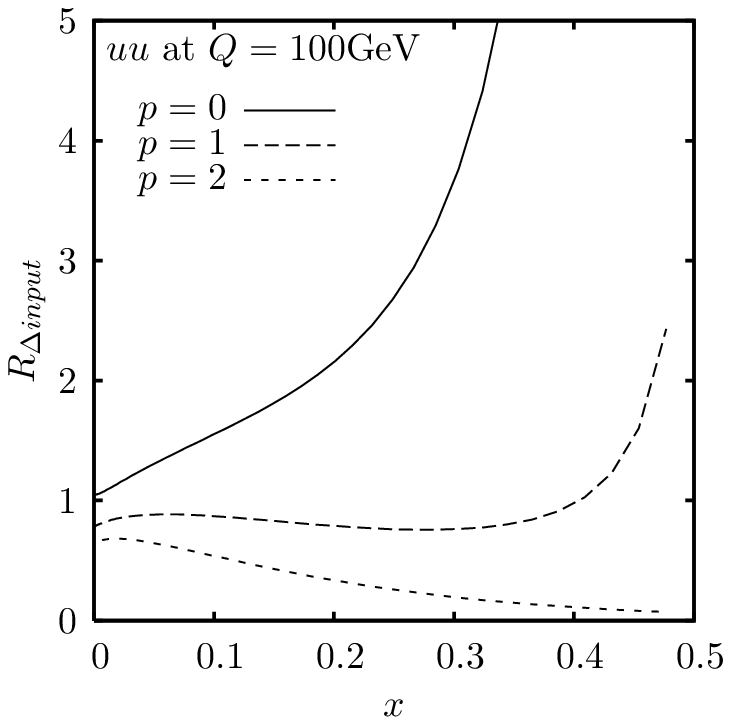}
\includegraphics[scale=0.9]{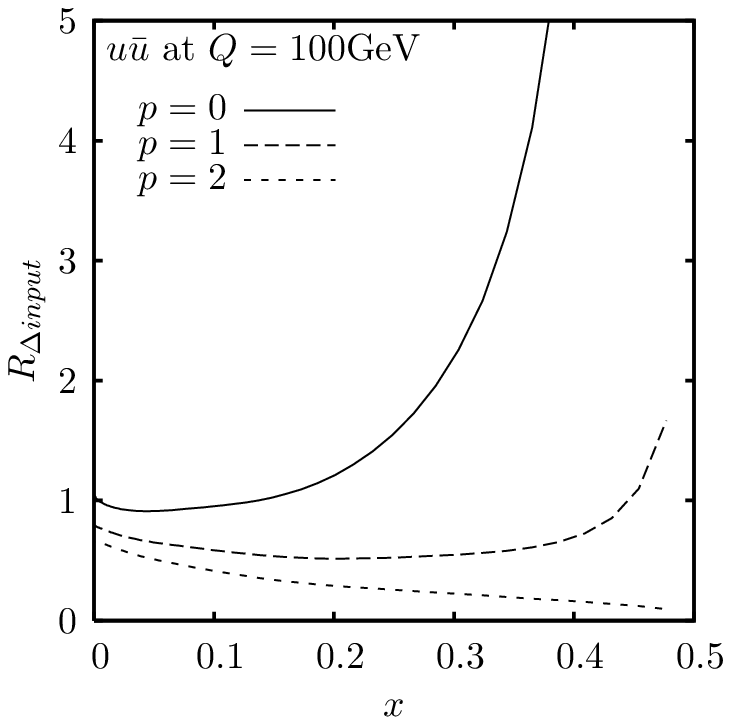}
\includegraphics[scale=0.9]{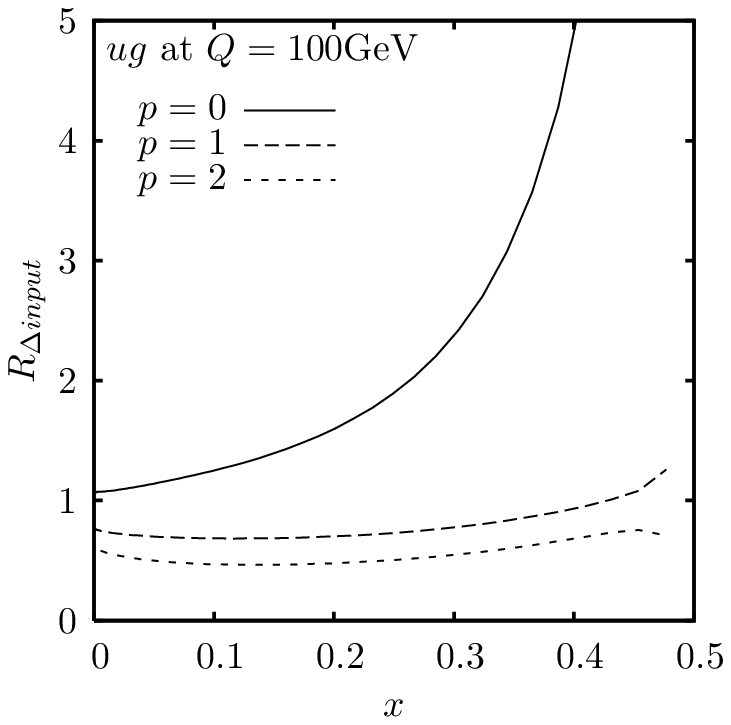}
\includegraphics[scale=0.9]{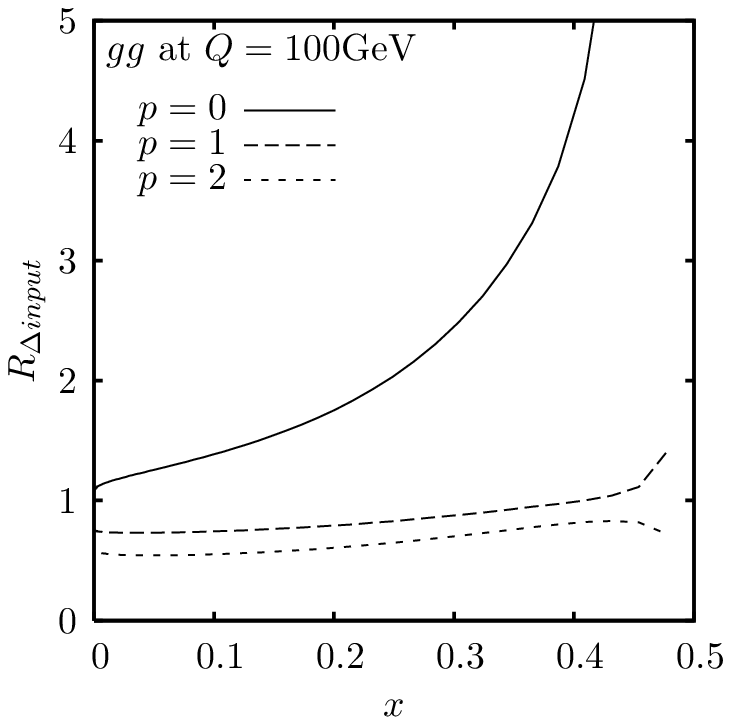}
\caption{\label{fig:DifferentInputs} Plots of the ratio $R^{ij}_{\Delta input}$ defined in equation \eqref{Rinputdef} for $Q=100$~GeV, $p=0,1$ and $2$, and the parton combinations $ij$ discussed in the text.}
\end{figure}

The main novel component of the present work is the introduction of the improved input dPDFs of Section \ref{sec:sumrulesinput}. Consequently, the first question we should like to answer is how use of the improved inputs in the dDGLAP equation, as opposed to naive `factorised$\times (1-x_1-x_2)^p$' inputs, affects the dPDFs at the scale $Q$. To this end, we have plotted the following ratio for our sample dPDFs in Fig.~\ref{fig:DifferentInputs}:
\begin{equation} \label{Rinputdef}
R^{ij}_{\Delta input}(x;Q) \equiv \dfrac{D_h^{ij}(x,x;Q) \mid_{ \text{input } D_h^{ij}(x_1,x_2;Q_0) = D_h^{i}(x_1;Q_0)D_h^{j}(x_2;Q_0)(1-x_1-x_2)^p}}
{D_h^{ij}(x,x;Q) \mid_{ \text{input } D_h^{ij}(x_1,x_2;Q_0) = \text{our improved inputs}}}
\end{equation}

We have made plots for each of the common traditional choices for $p$ -- $0,1$ and $2$. One immediately notices in Fig.~\ref{fig:DifferentInputs} that all of the ratio curves deviate significantly from $1$. This shows that the precise choice of inputs at the low scale has an important impact on the high scale dPDFs, and demonstrates the inadequacy of the traditional naive input forms. We see that multiplying factorised inputs by a phase space factor of $(1-x_1-x_2)$ or $(1-x_1-x_2)^2$ gives high scale dPDFs which are generally too small for small $(x_1,x_2)$. This is expected -- we have seen that $(1-x_1-x_2)$ or $(1-x_1-x_2)^2$ phase space factors suppress the inputs too much in the high $x_1$, low $x_2$ and high $x_2$, low $x_1$ regions. Since these regions directly feed the small $x_1,x_2$ region, this directly translates into a deficiency in the high scale dPDFs in the small $x_1,x_2$ region. Conversely, we see that not using a phase factor in the inputs results in high scale dPDFs which are generally too large. This is because in this scenario the inputs are too large near the kinematic bound, and this excess propagates down to smaller $x_1,x_2$ values during evolution.

It is interesting to note that, contrary to the previous general statement, the $p=0$ ratio 
for the $u\bar{u}$ dPDF actually dips below unity between $x=0.005$ and $x=0.15$. Furthermore, we see that 
the $p=0$ $uu$ ratio rises above the corresponding ratios for the other flavour combinations. 
The origin of each of these features is in the extra terms we included in our
improved inputs to take account of valence number effects or $j\bar{j}$ correlations, 
which do not appear in the naive inputs. The inclusion of a positive $j\bar{j}$ correlation term in the
 $u\bar{u}$ distribution causes our $u\bar{u}$ dPDF to be larger at the high scale than 
it would be if the correlation term were absent. Since our dPDFs appear on the denominator of $R^{ij}_{\Delta input}$,
 this manifests itself as a reduction in our $p=0$ $u\bar{u}$ ratio. Conversely, the 
subtraction of a valence number effect term from our $uu$ input results in a reduction of 
our $uu$ dPDF at $Q$, which increases the $uu$ ratio.

For $p=1$ and $2$, we observe that the $uu$ ratio is still larger than the others for small $x$. 
However, the $u\bar{u}$ ratio is now very slightly larger than the $ug$ and $gg$ ratios 
at small $x$ values. This is because the $ug$ and $gg$ high scale distributions at small 
$x$ are more sensitive to the form of the input distributions near the kinematic boundary than 
the $u\bar{u}$. This is a simple consequence of the fact that gluon type evolution causes a 
faster cascade of PDFs to low $x$ values than $u$ or $\bar{u}$ type evolution. The reduction 
in the $ug$ and $gg$ ratios at small $x$ relative to the $u\bar{u}$ due to the change in $p$ 
overcomes the small effect of including the $j\bar{j}$ correlation term in our $u\bar{u}$.

The contributions of the $j\bar{j}$ correlation and valence number effect terms 
to the high scale ($Q=100$~GeV) double human basis dPDFs are most cleanly observed at $x \sim 0.05$, 
and are on the order of $10\%$ in this $x$ region. For smaller $x$, the contributions 
from the extra terms are swamped by sea-sea contributions to the dPDF, whilst at larger $x$, 
phase space effects become dominant.

Aside from looking at the effect of using different inputs on the dPDFs at scale $Q$, we can also ask to what extent correlations introduced by dDGLAP evolution affect the dPDFs at $Q$.  There are essentially two types of correlations that the dDGLAP equation introduces -- correlations due to the requirement of momentum conservation, and more interesting correlations generated by the sPDF feed terms. Here, we choose to look specifically at the effect of the latter.

In order to do this, we evolved our improved input dPDFs up to the scales $Q=10$~GeV, $Q=100$~GeV, and $Q=1000$~GeV, both with the sPDF feed terms included in the evolution, and also with these terms set to zero. For each final scale and parton pairing in our selected set, the following ratio was then plotted:
\begin{equation} \label{Rfeeddef}
R^{ij}_{\text{no feed}}(x;Q) \equiv \dfrac
{D_h^{ij}(x,x;Q) \mid_{\text{our improved inputs, no sPDF feed}}}
{D_h^{ij}(x,x;Q) \mid_{\text{our improved inputs}}}
\end{equation}

\begin{figure}
\centering
\includegraphics[scale=0.9]{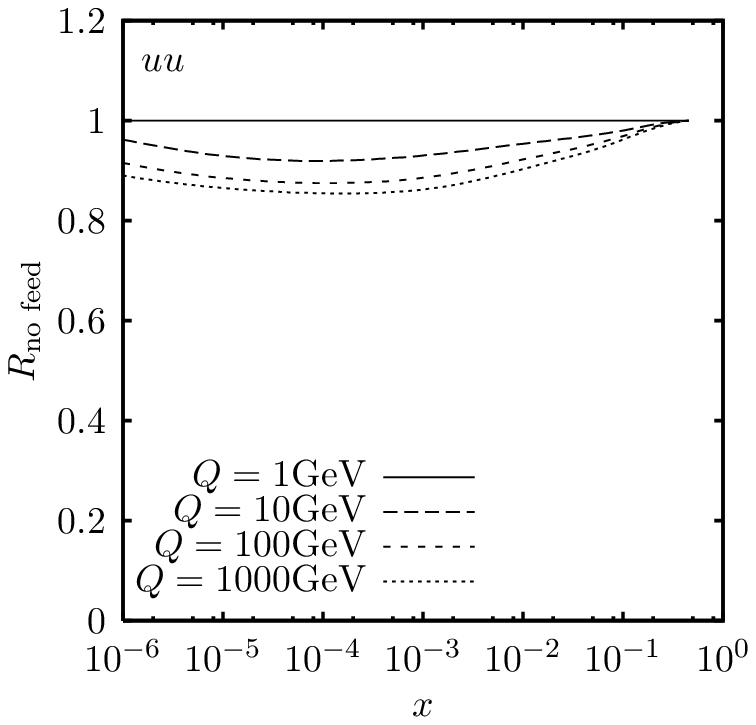}
\includegraphics[scale=0.9]{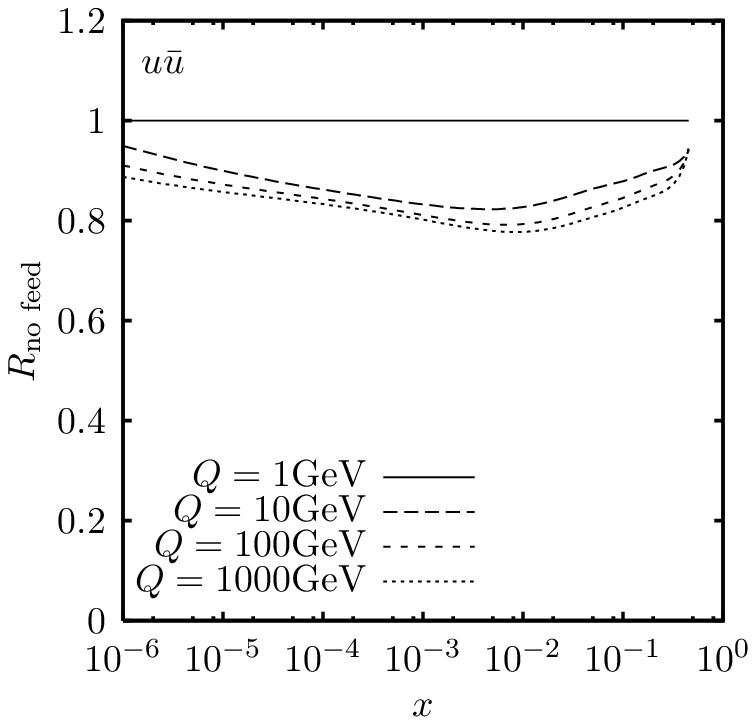}
\includegraphics[scale=0.9]{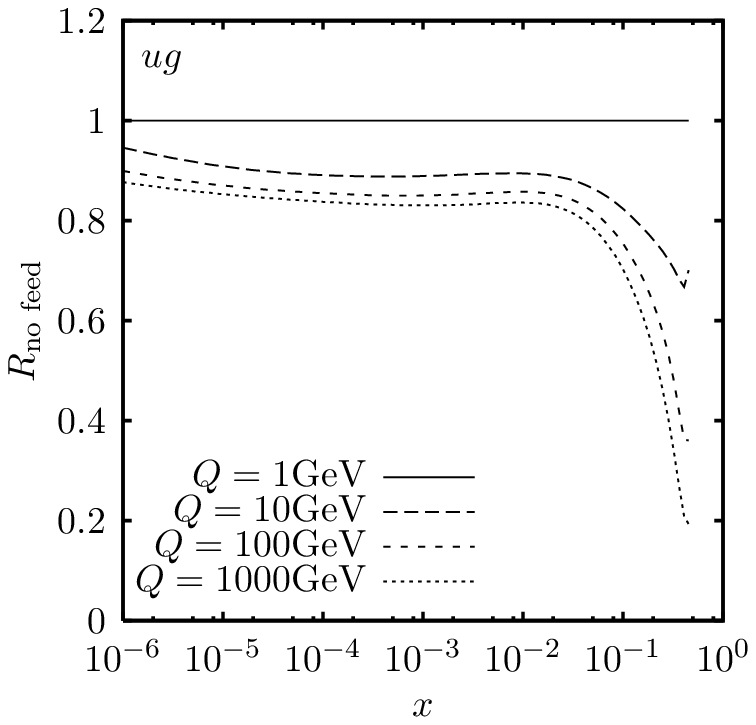}
\includegraphics[scale=0.9]{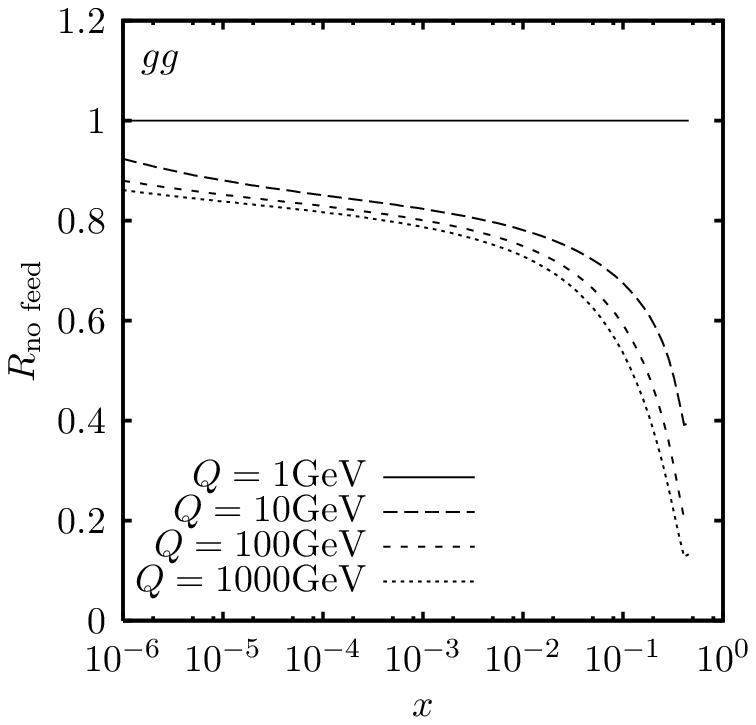}
\caption{\label{fig:MinusFeed} Plots of the ratio $R^{ij}_{\text{no feed}}$ defined in equation \eqref{Rfeeddef} for $Q=1,10,100$ and $1000$~GeV and the parton combinations $ij$ discussed in the text.}
\end{figure}

\begin{figure}
\centering
\includegraphics[scale=1]{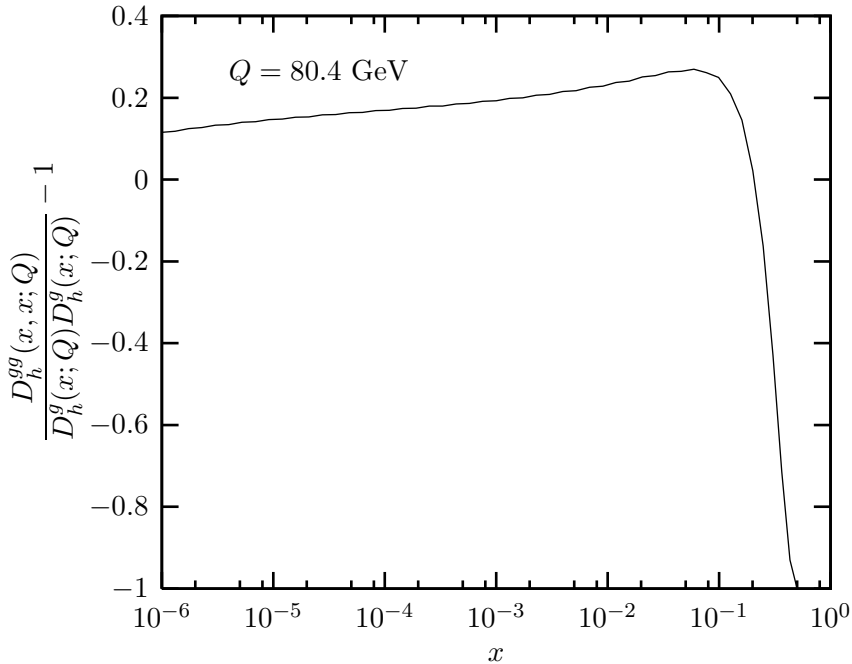}
\caption{\label{fig:ggratio} $gg$ correlation ratio $R^{gg}$ at $Q=80.4$~GeV obtained using MSTW2008LO factorised inputs.}
\end{figure}

 We plot the results using a logarithmic $x$ scale in Fig.~\ref{fig:MinusFeed} \footnote{In this figure, and in figures \ref{fig:ggratio} and \ref{fig:ggrays}, we make plots down to $x = 10^{-6}$. Although it is interesting to look at our LO dPDFs at very small $x$, we should mention that we do not expect the leading order approximation to produce very accurate dPDFs in this region.}.  The effect of the sPDF terms is small but non-negligible, being at roughly the $10\%$ level for $x<10^{-2}$ in all of the dPDFs considered, and increasing with $Q$. 

We observe that the ratios for all of the given flavour combinations look very similar for $x$ from $10^{-6}$ to $10^{-4}$. The reason for this is that the small $x$ shape of the distributions considered is very strongly determined by the (either direct or indirect) feeding of these distributions by the $gg$ distribution. If the $gg$ dPDF loses its sPDF feed and is reduced by a certain percentage at small $x$, the connection of the other dPDFs to the $gg$ will result in these dPDFs being reduced by a similar amount. This explanation can be verified by investigating what happens if we remove all of the sPDF terms except for the $gg$ feed. In this case the ratios for all of the considered dPDFs are much closer to $1$ for $10^{-6}<x<10^{-4}$, suggesting that the subtraction of the $gg$ sPDF feed is the dominant factor determining the shapes of the plots in Fig.~\ref{fig:MinusFeed} for small $x$.

For larger $x$, the deviation of the $uu$ ratio from $1$ remains small, and tends to $0$ as $x$ approaches its maximum of $0.5$. This is expected since there is no direct sPDF feed term in the evolution of the $uu$ dPDF. The $u\bar{u}$ ratio also seems to tend to $1$ as $x \to 0.5$, albeit more slowly, whilst the $ug$ and $gg$ ratios plunge towards zero, the $gg$ more rapidly than the $ug$. This implies that at large $x$, the sPDF feed contributions are more important to the $gg$ than they are to the $ug$, and that they are more important to the $ug$ than they are to the $u\bar{u}$. We can explain this ordering using a fact we have previously mentioned -- namely, that the `pull' on a gluon PDF towards lower $x$ values during evolution is stronger than that on a quark type PDF. The $gg$ distribution at large $x$ is pulled strongly towards lower $x$ values in two directions, and is very much smaller if it is not continuously fed by an sPDF. By contrast, the `pull' on the large $x$ $u\bar{u}$ distribution is smaller in both directions, and so the contribution of similar sPDF feed terms is proportionately smaller. The $ug$ distribution has one gluon flavour index and one quark, so the importance of the sPDF feed on this distribution at large $x$ is intermediate.

We have not been able to exactly reproduce the results of either of the extant numerical investigations into the correlations induced by evolution -- \cite{Cattaruzza:2005nu} and \cite{Korotkikh:2004bz}. However, we do agree with \cite{Korotkikh:2004bz} that the accumulated
sPDF feed contribution to the $gg$ between $\sim 1$~GeV and $100$~GeV accounts for about $10\%$ of the $Q=100$~GeV $gg$ distribution at small $x$.  In Fig.~\ref{fig:ggratio}, we plot the following ratio for $Q=80.4$~GeV:
\begin{equation} \label{Rggdef}
R^{gg}(x;Q) \equiv \dfrac
{D_h^{gg}(x,x;Q) \mid_{ \text{factorised inputs}}-D_h^{g}(x;Q)D_h^{g}(x;Q)}
{D_h^{g}(x;Q)D_h^{g}(x;Q)}
\end{equation}

This figure corresponds to the solid curve in Fig.~1 of \cite{Cattaruzza:2005nu}, with MSTW2008LO inputs replacing the MRS99 inputs used there. We expect that the ratio $R^{gg}$ should tend to $-1$ as $x$ approaches $0.5$ for any $Q$ sufficiently larger than the input scale. This is because evolution will very quickly cause $D_h^{gg}$ to become much smaller than the factorised value near the kinematic bound. Our curve exhibits this property, but it seems unlikely that the solid curve plotted in Fig.~1 of \cite{Cattaruzza:2005nu} will, especially if it reaches $0.6$ for higher $x$ values as is stated in \cite{Cattaruzza:2005nu}.

\begin{figure}
\centering
\includegraphics[scale=0.9]{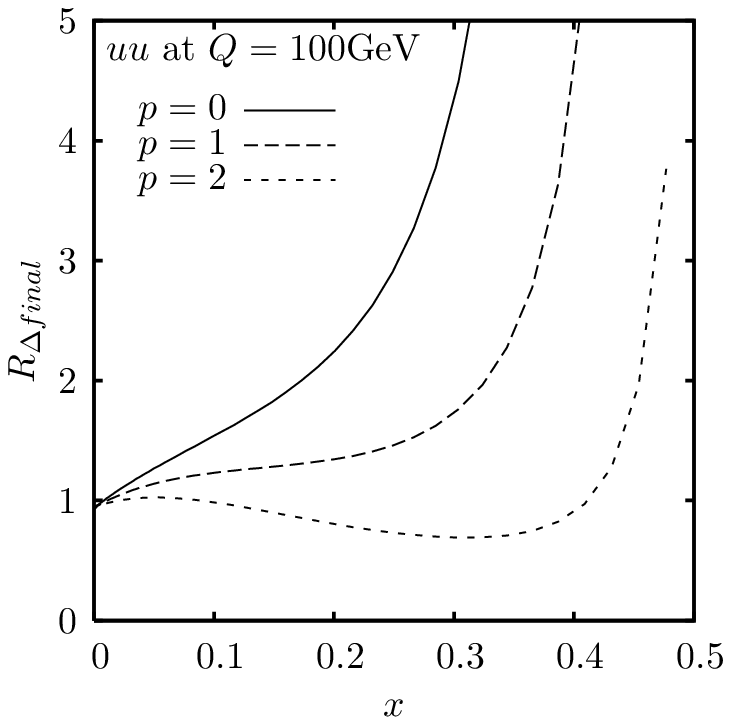}
\includegraphics[scale=0.9]{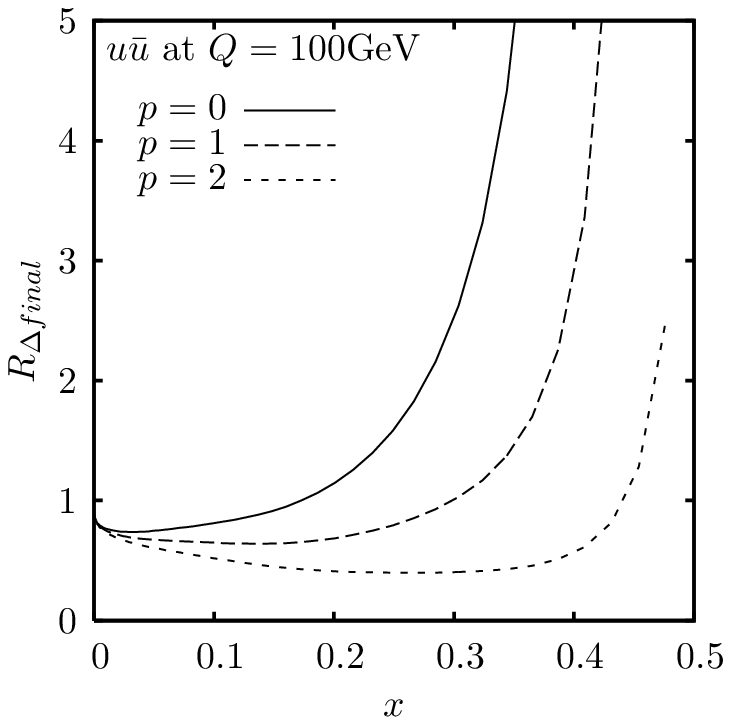}
\includegraphics[scale=0.9]{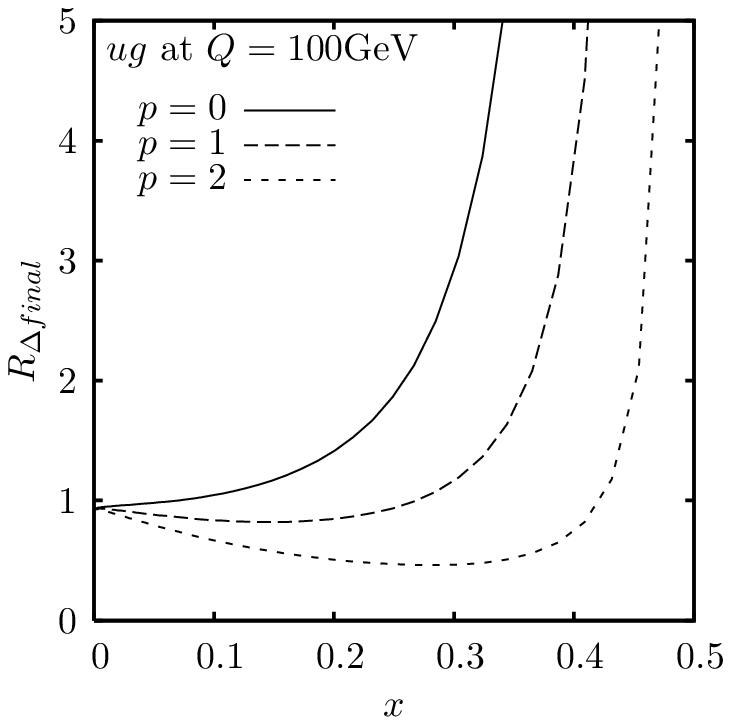}
\includegraphics[scale=0.9]{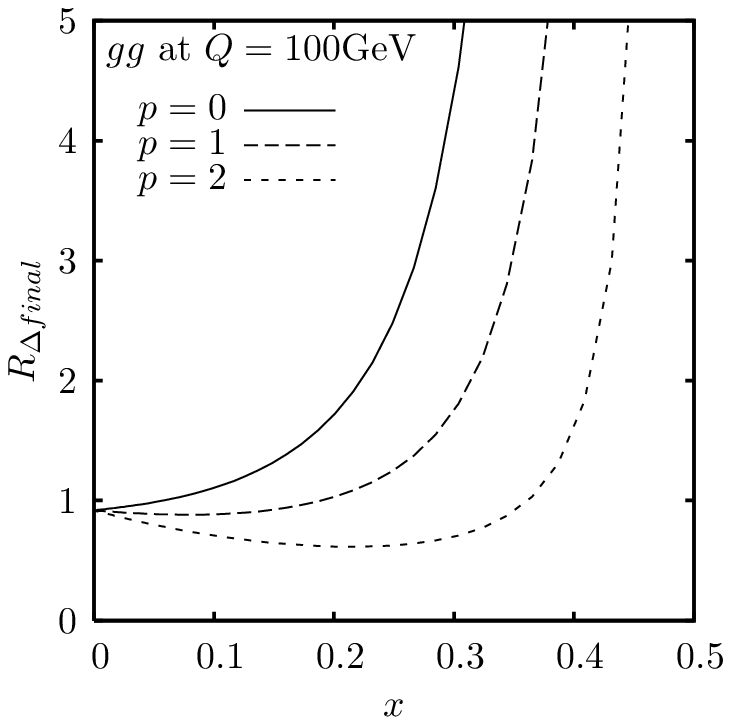}
\caption{\label{fig:DifferentFinals} Plots of the ratio $R^{ij}_{\Delta final}$ defined in equation \eqref{Rfinaldef} at $Q=100$~GeV and along the line $x_1=x_2=x$. The ratio is plotted for $p=0,1$ and $2$ and for each of the parton combinations $ij$ discussed in the text.}
\end{figure}

\begin{figure}
\centering
\includegraphics[trim = 1.9cm 0 0 0, scale=1]{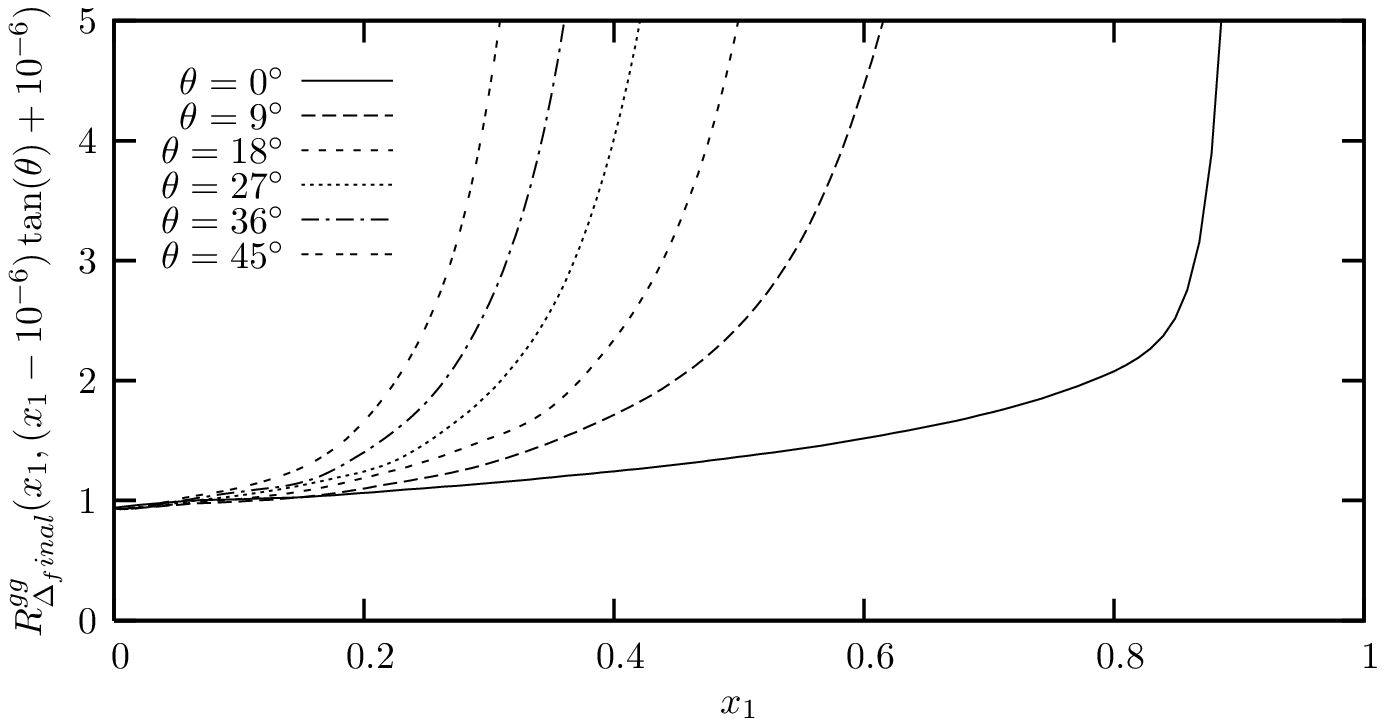}
\caption{\label{fig:ggrays} The ratio $R^{gg}_{\Delta final}$ plotted along various lines of the form $x_2=(x_1-10^{-6})\tan(\theta)+10^{-6}$ at $Q=100$~GeV.}
\end{figure}

Finally, we compare our full treatment (improved inputs plus full dDGLAP evolution) with the approximation that simply uses factorised inputs $\times (1-x_1-x_2)^p$ ($p = 0,1$ or $2$) at the scale $Q$. This approximation is frequently used in phenomenological studies of double parton scattering processes. In Fig.~\ref{fig:DifferentFinals}, we plot the following ratio along the line $x_1=x_2=x$ for our sample dPDFs and for $p = 0,1$ and $2$:
\begin{equation} \label{Rfinaldef}
R^{ij}_{\Delta final}(x_1,x_2;Q) \equiv \dfrac{D_h^{i}(x_1;Q) D_h^{j}(x_2;Q)(1-x_1-x_2)^p}
{D_h^{ij}(x_1,x_2;Q) \mid_{\text{our improved inputs}}}
\end{equation}

The plots reveal that even a $(1-x_1-x_2)^2$ phase space factor multiplying a factorised form at $Q$ underestimates the large $x$ falloff in the dPDFs along $x_1=x_2=x$. For very small $x$, the ratios are all slightly less than $1$ due to the fact that one misses the sPDF feed contributions if one uses a factorised form at $Q$ (note that the ratio appears smallest at very low $x$ for the $u\bar{u}$, due to the fact that the sPDF feed for the $u\bar{u}$ is particularly important around $x=10^{-2}$ -- see Fig.~\ref{fig:MinusFeed}). One also notices the imprint of omitting the valence number effect and $j\bar{j}$ correlation terms in the ratios -- the $uu$ ratio rises above the others at $x\sim 0.05$, whilst the $u\bar{u}$ dips at this $x$ value.

It is interesting to consider the behaviour of $R^{ij}_{\Delta final}(x_1,x_2;Q)$ away from
the line $x_1=x_2=x$. In Fig.~\ref{fig:ggrays}, we plot the $p=0$ ratio for the $gg$ flavour combination along several lines emanating from the point $x_1=10^{-6}, x_2=10^{-6}$. The 
figure shows that the deviation of this ratio from $1$ is maximal along $x_1=x_2$ (in fact, this statement holds for any combination of parton indices). We observe that a $p=0$ factorised form is a fairly good approximation to our $gg$ dPDF close to the $x_1$ axis, except when $x_1$ is very large ($x_1>0.8$). This is to be expected, given our use of input dPDFs which essentially reduce to $p=0$ factorised forms near the lines $x_1=0$ and $x_2=0$. One can infer from the plot that use of a factorised form multiplied by either $(1-x_1-x_2)$ or $(1-x_1-x_2)^2$ will result in one overestimating the falloff in the dPDFs in the $x_1 \sim 0, x_2 \lesssim 0.8$ and $x_1 \lesssim 0.8, x_2 \sim 0$ regions.

\section{Summary and Outlook} \label{sec:conclusion}

In this report, we have developed a framework based on the dDGLAP equation for
calculating the LO double distributions $D_h^{ij}(x_1,x_2;t)$ which represents
an improvement on approaches used previously. We have derived for the first
time the momentum and sum rules that the dPDFs have to obey. An important
implication of these sum rules is that the conventionally held wisdom that the
dPDFs should be approximately equal to products of sPDFs for small $x_1,x_2$
does not apply in the case of the equal flavour valence-valence dPDFs. Using
the dPDF sum rules, we have constructed a set of improved input dPDFs corresponding to
the MSTW2008LO sPDF inputs. In the double human flavour basis, these dPDFs are 
all positive and satisfy the sum rules to better than $25\%$ precision (for $x<0.8$).

We have written a program which numerically integrates the LO dDGLAP equation
using a direct $x$ space method, enabling one to evolve the dPDF inputs to 
higher scales. The accuracy of the program is good for small $x_1,x_2$ -- an evolution from
$1$~GeV to $100$~GeV using a grid with only $150$ points in each $x$ direction
and $10$ points in the $t$ direction produces dPDF values with numerical errors of 
less than $1\%$ for $x_1<0.3, x_2<0.3$. We have produced a set of publicly available
dPDF grids by applying the numerical procedure to our improved inputs, which can be found along with  
interpolation code at \cite{HepForgePage}. To produce the grids, $600$ points were used in each
$x$ direction, and $120$ in the $t$, ensuring an accuracy much better than $1\%$ for small $x$.

We saw that the accuracy of our program is rather poor near the kinematical
bound. If the accuracy here needed to be improved without significantly
increasing computing time, then a multigrid method could be implemented in the
program (for an example of the use of this method for the sPDF case, see
\cite{Botje:2009bj}). The additional more finely spaced grids would be
introduced in the region near the kinematic bound to increase accuracy in this
region.

For the purposes of experimental studies of double parton interactions in the 
near future, which will be attempting to establish the existence of correlations
in the dPDFs, the LO treatment presented here is sufficiently accurate. If 
correlations are found and they agree with some or all of the predictions made 
here, then this will be a strong impetus for us to extend the formalism to NLO.
As mentioned in Section~\ref{sec:theory}, such an extension is not trivial, since 
the structure of the third term of the dDGLAP equation becomes significantly more
complex at NLO. At this order, one requires the functions $P_{i \to jk}^{(1)}(x_1,x_2)$ which cannot be obtained trivially from the NLO sPDFs as in the LO case. It is likely
that these functions exist in the literature, although some work may need to be done
to get them into a form that can be used in the dDGLAP equation. 

Many double scattering processes which might provide important
signals/backgrounds at the LHC do not involve the same hard scale in both
collisions. An example of such a process is the simultaneous production of a $W$ and a
$b\bar{b}$ pair in separate collisions. This forms a background to the process
$p+p \to WH$, $H \to b\bar{b}$, which might be an important process to discover
the Higgs if $m_h<2m_W$ \cite{DelFabbro:1999tf}. To make theoretical predictions 
relating to these processes, we require the more general double distributions with 
$t_1 \ne t_2$. As is mentioned in Section \ref{sec:theory}, we believe that these 
distributions are calculated by adding an extra sDGLAP evolution in one variable on
top of the dDGLAP evolution. A useful extension to the work would be to produce a
more general set of double distributions based on this hypothesis.

Finally, there exists the possibility of using the dPDFs developed above to
undertake a phenomenological investigation of double parton scattering at the
LHC. In particular it would be interesting to examine how the `correlations'
introduced via our inputs and by evolution affect the properties of a
double scattering event, and also how one might measure the correlations in
practice. We are currently in the process of making such an investigation. 

\section*{Acknowledgements}

JG acknowledges financial support from the UK Science and Technology Facilities Council.

\newpage

\section*{Appendix}

\appendix
\section{Numerical techniques for evaluating the dDGLAP integrals} \label{app:NumInt}

Let us consider the integrals which have to be numerically approximated using
the $(x_1,x_2)$ grid. All of these integrals are of the following schematic
form:

\begin{equation} \label{schemint}
I(y) = \int_{x}^{1-y}\dfrac{dz}{z}D(z,y)P\left(\dfrac{x}{z}\right)
\end{equation}

The splitting function $P(x)$ may in general consist of three terms. The first
of these is a regular term $A(x)$ and the second is a term proportional to a
delta function $K \delta(1-x)$. The final term consists of a product of two
factors. The first of these is a simple regular function $R(x)$, whilst the
second is a function $S(x)$ containing a singular factor $1/(1-x)$ which is
regularised by the plus prescription:
\begin{equation} \label{schemsplit}
P(x) = A(x) + K \delta(1-x) + R(x)[S(x)]_{+} .
\end{equation}
Inserting the form \eqref{schemsplit} into \eqref{schemint}, we find that the
integrals which have to be approximated using the grid have the following
general form:
\begin{eqnarray} \label{genint}
I(y)  &=& I_1(y)+I_2(y)+I_3(y) \hspace{5pt} \text{with}\\ \label{genint1}
I_1(y)&\equiv &\int_{x}^{1-y}\dfrac{dz}{z}D(z,y)A\left(\dfrac{x}{z}\right)
\\ \label{genint2}
I_2(y)&\equiv &KD(x,y)
\\ \nonumber
I_3(y)&\equiv & \int_x^{1-y}\dfrac{dz}{z}S\left(\dfrac{x}{z}\right)
\left[D(z,y)R\left(\dfrac{x}{z}\right)-\dfrac{x}{z}D(x,y)R(1)\right]
\\ \label{genint3}
&-&R(1)D(x,y)\int_0^{x/(1-y)}dzS(z) .
\end{eqnarray}
The integral in the last term of \eqref{genint3} can be done analytically for
each splitting function. The integrals in \eqref{genint1} and the first term of
\eqref{genint3} are the ones that must be performed on the grid. We note that
the integrand in the first term of \eqref{genint3} has the property that it is
undefined for $z=x$ (due to the fact that $S(x/z)$ contains a factor
$1/(1-x/z)$). It nevertheless tends to a finite limit as $z \to x$ (due to the
fact that the divergence in $S(x/z)$ is compensated for by the other factor in
the integrand going to zero as $z \to x$). This suggests the use of a method
for performing the numerical integrations which effectively estimates the
integrand between $z=x$ and the grid point with next highest $z$ by
extrapolating from integrand values on nearby grid points (with higher $z$).

A method which uses an open Newton-Cotes rule of degree $n$ for the first $n$
integration intervals, and then switches to a closed Newton-Cotes rule to
perform the integration over the remaining intervals, has this property. If the
number of integration intervals is greater than $3$, we use Simpson's rule as
the closed rule, combined with an open rule of degree $4$ when the number of
integration intervals is even, and an open rule of degree $5$ otherwise. Open
rules of the appropriate degree are used on their own when the number of
intervals is $3$ or fewer. This ensures an overall integration method which for
most integrals has an error of $O(n\Delta u^5\tfrac{df^{(4)}(\xi)}{du^4})$. In
this formula, $n$ is the number of intervals used, $\Delta u$ is the (even)
grid spacing in $u=\ln(\tfrac{x}{1-x})$, $f$ is the integrand taking into
account the Jacobian on the transformation into $u$ space, and $\xi$ is the
value of $u$ that maximises $df^{(4)}/{du^4}$.

With the numerical method described, the integral \eqref{schemint} is
approximated by:
\begin{eqnarray} \label{approxint}
I(y) &\approx& \sum\limits_{j=i+1}^k
D(z_j,y)\left[A\left(\dfrac{z_i}{z_j}\right) +
R\left(\dfrac{z_i}{z_j}\right) S\left(\dfrac{z_i}{z_j}\right)
\right]w_{ijk}\dfrac{J(z_j)}{z_j}\Delta u  \nonumber \\
&+& D(z_i,y)\biggl[K-R(1)\int_0^{x/(1-y)}dzS(z)  \nonumber \\
&-&\sum\limits_{j=i+1}^kS\left(\dfrac{z_i}{z_j}\right)\dfrac{z_i}{z_j}R(1)
w_{ijk}\dfrac{J(z_j)}{z_j}\Delta u\biggr] .
\end{eqnarray}
The indices $\{i,j,k\}$ represent grid points, with $i$ corresponding to the
grid point with $z$ value equal to $x$ ($z_i \equiv x$) and $k$ corresponding
to the point with $z$ value equal to $1-y$ ($z_k \equiv 1-y$). The $w_{ijk}$
are Newton-Cotes type integration weights whose values are dictated by the
prescription described above. Note that the weight at grid point $j$ under this
prescription depends on the start and end points of the integration -- hence $w$
depends on the indices $i$ and $k$. The function $J(x)$ is the Jacobian, $J(x)
\equiv dx/du = x(1-x)$.

We may rewrite \eqref{approxint} as:
\begin{equation}
I(y) \approx \sum\limits_{j=i}^kP_{ijk}D(x_j,y) ,
\end{equation}
where
\begin{eqnarray}
P_{ijk} = \begin{cases}
\left[A\left(\dfrac{z_i}{z_j}\right) +
R\left(\dfrac{z_i}{z_j}\right) S\left(\dfrac{z_i}{z_j}\right)
\right]w_{ijk}\dfrac{J(z_j)}{z_j}\Delta u & \text{if $i<j\le k$}
\vspace{10pt}\\
K-R(1)\int_0^{x/(1-y)}dzS(z)
\\-\sum\limits_{j=i+1}^kS\left(\dfrac{z_i}{z_j}\right)\dfrac{z_i}{z_j}R(1)
w_{ijk}\dfrac{J(z_j)}{z_j}\Delta u & \text{if $j=i$, $i<k$} \vspace{10pt} \\
0 & \text{otherwise.}
\end{cases}
\end{eqnarray}
The three-dimensional array $P_{ijk}$ only depends on the splitting function
$P(x)$, Jacobian $J(x)$ and weights $w_{ijk}$. None of these vary during an
evolution, with the possible exception of $P_{gg}$ (this contains a term
proportional to $n_f$ in the $K\delta(1-x)$ piece and so may vary in a variable
flavour number scheme -- see Section~\ref{subsec:flavour}). We therefore precalculate and store the
elements of $P_{ijk}$ during program initialisation, to increase efficiency.
The possible variation of the contributions to $P_{ijk}$ from the term in
$P_{gg}$ proportional to $n_f$ is handled by postponing the calculation of
these contributions such that they are calculated and reintroduced at each
evolution step (using the value of $n_f$ appropriate to that step).

\newpage

\bibliography{DPFpaper_v1}{}
\bibliographystyle{JHEP}

\end{document}